# Super-resolution imaging reveals resistance to mass transfer in functionalized stationary phases

Ricardo Monge Neria,[1] Muhammad Zeeshan,[2] Aman Kapoor,[1] Tae Kyong John Kim,[3] Nichole Hoven,[3] Jeffrey S. Pigott,[3] Burcu Gurkan,[2] Christine E. Duval,[2] Rachel A. Saylor,[5] Lydia Kisley[1,4]*

[1]Department of Physics, [2]Department of Chemical and Biomolecular Engineering, [3]Swagelok Center for Surface Analysis of Materials, and [4]Department of Chemistry, Case Western Reserve University, Cleveland, Ohio 44106-7079
[5]Department of Chemistry and Biochemistry, Oberlin College, Oberlin, OH 44074
*lydia.kisley@case.edu, ORCID: 0000-0002-8286-5264

## One Sentence Summary

Nanoscale *in situ* imaging and chromatography show chemical separations can be improved by reducing polymers that block pores.

## Abstract

Chemical separations are costly in terms of energy, time, and money. Separation methods are optimized with inefficient trial-and-error approaches that lack insight into the molecular dynamics that lead to the success or failure of a separation and, hence, ways to improve the process. We perform super-resolution imaging of fluorescent analytes in four different commercial liquid chromatography materials. Surprisingly, we observe that chemical functionalization can block over 50% of the material's porous interior, rendering it inaccessible to small molecule analytes. Only *in situ* imaging unveils the inaccessibility when compared to the industry-accepted *ex situ* characterization methods. Selectively removing some of the functionalization with solvent restores pore access without significantly altering the single-molecule kinetics that underlie the separation and agree with bulk chromatography measurements. Our molecular results determine that commercial "fully porous" stationary phases are over-functionalized and provide a new avenue to characterize and direct separation material design from the bottom-up.

## Main Text

The separation of molecules from mixtures uses ~15% of the total energy in the U.S. (*1*), emits 100 million tons of $CO_2$, and costs $4 billion annually (*2*). Many separation processes involve the use of chromatography, which relies on differences in molecular-scale interactions of analytes with a stationary phase to achieve separation. Yet, the development of chromatographic separations is done through intensive, iterative, trial-and-error "top-down" approaches. Such ensemble-averaged experimental and theoretical (*3–5*) methods obscure the mass transfer and molecular interactions that lead to either the failure or success of a separation. A "bottom-up" approach resolving the molecular behavior and local physical phenomena provides insight into ways to address challenges in separations, aiding the development of new technologies (*1, 6*).

Specifically, for liquid chromatography − where an analyte in a liquid mobile phase differentially interacts with a solid stationary phase compared to other molecules for separation – current characterization of stationary phase materials is performed *ex situ*. Common approaches require conditions that are far removed from the liquid environment in which the separation occurs. Nitrogen adsorption isotherms measure gas uptake in mg-g's of stationary phase under non-hydrated conditions, low 77 K temperatures, and use an assumed monolayer theory to extract surface area, pore volumes, and sizes (*7–9*). Inverse size-exclusion chromatography of various-sized probe molecules, such as dextran, indirectly measure pore accessibility under hydrated conditions, but average behavior over ≥$10^{10}$ stationary phase particles (*5*), assume uniform pore shape, and are limited to standard probes that obscure the chemical complexities of real analytes (*9–11*). Nanoscale electron microscopy provides direct spatial measurements of pore sizes and morphology, but requires vacuum operating conditions and destructive cross sectioning to image the internal structure of materials (*9, 12*).

In this work, we improve upon and apply our single-molecule light-sheet super-resolution imaging method (*13*) to characterize chromatographic materials at the nanoscale under *in situ* conditions relevant to real liquid chromatography applications. We simultaneously super-resolve both the location and adsorption kinetics of single-molecule analytes within four different commercial stationary phases (Fig. 1A, table S1). Our imaging spatially reveals limited accessibility to the inner volume of the functionalized chromatography particles commercially sold as "fully porous." Access can be regained by decreasing the degree of functionalization without affecting the adsorption kinetics that underly the bulk elution behavior in high performance liquid chromatography (HPLC). The results question previous assumptions about the understanding of mass-transfer behavior at nano- and microscales within stationary phases.

**The inner volume of porous chromatography particles functionalized with cellulose is inaccessible to analytes**

We study stationary phase materials that vary in porosity and chemical functionalization (Fig. 1A, table S1). Our main material of interest is Cellulose-B, marketed as a fully porous particle (FPP) used for chiral separations (*14, 15*). Cellulose-B is sold as 5 µm diameter silica particles, with 100 nm wide pores, functionalized with cellulose tris(3,5 dimethyl phenyl carbamate). To understand the effect of functionalization, Cellulose-B is compared to FPPs of bare silica of the same porosity and size. We also image superficially porous particles (SPP) comprised of a 1.7 µm diameter solid core surrounded by a 0.5 µm thick, 10 nm porous silica shell, functionalized with 1-(3,5-dinitrobenzamido)-1,2,3,4,-tetrahydrophenanthrene (Whelk-O1). The Whelk-O1 SPPs are only available at a 2.7 µm diameter and subsequent analysis is normalized to account particle size. We also characterize FPPs with Whelk-O1 functionalization to support our results (fig. S2).

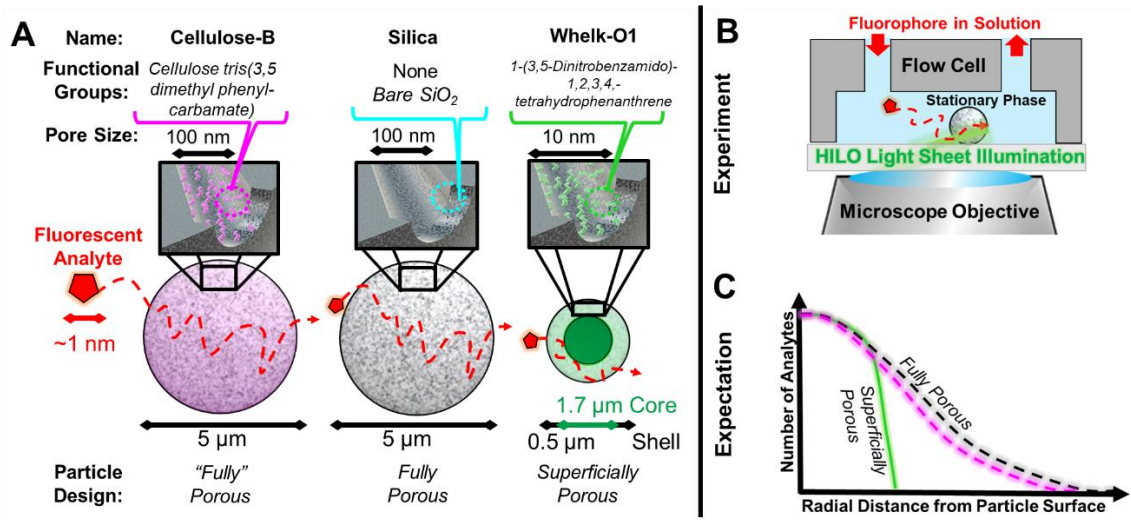

**Fig. 1. Supplier-provided properties of commercial stationary phase particles studied by super-resolution imaging.** (**A**) Commercial porous stationary particles tested in this study. We image 5 µm diameter Cellulose-B functionalized (magenta) and bare non-functionalized silica particles (grey) FPPs, both with 100 nm pores. We further test Whelk-O1 functionalized particles (green) with 10 nm pores and a SPP design that has a 1.7 µm diameter non-porous silica core. We use rhodamine 6G (red) as a fluorescent analyte that can diffuse and adsorb throughout the porous stationary phases. (**B**) The fluorophores are excited and imaged diffusing in solution and in the stationary phases by single-molecule HILO fluorescence microscopy. (**C**) Analytes are expected to freely diffuse throughout the full volume of the FPPs but should be blocked from accessing the center of SPPs due to the solid core. See table S1 for further details.

We dynamically super-resolve the locations of individual fluorescent analyte molecules as they diffuse and adsorb within chromatography particles (Fig. 1B). Rhodamine 6G is our model analyte; a ~1 nm wide molecule with a measured hydrodynamic radius of 0.589 nm (*16*, *17*), being at least 100 times smaller than the Cellulose-B and silica pores (100 nm) and comparable in size to small molecules separated on these columns (*18–20*). Analytes are imaged with our Highly Inclined and Laminated Optical (HILO) sheet microscopy, which we improve upon by reducing the sheet thickness to 2.3 ± 0.5 µm (fig. S1) (*13*). Analytes are expected to fully diffuse throughout the volume of the Cellulose-B and silica FPPs, while being blocked by the solid core in Whelk-O1 SPPs (Fig. 1C). Diffraction-limited fluorescence imaging of single molecules (Fig. 2A) is enhanced by applying localization analysis, where individual adsorbed molecules are identified with 32 ms temporal resolution and 30 ± 20 nm lateral and 340 ± 50 nm axial resolutions (*13*). The localizations produce super-resolution images of analyte adsorption sites within stationary phase particles (Fig. 2B, 2C), obtaining nanoscale information that would be otherwise obscured in diffraction-limited imaging (*21*, *22*). Images were obtained throughout the ~10-65 µm³ 3D volumes of the particles (Figs. 2D, 2F, 2H), with individual, 2D slices in 250 ± 10 nm steps (Figs. 2E, 2G, 2I).

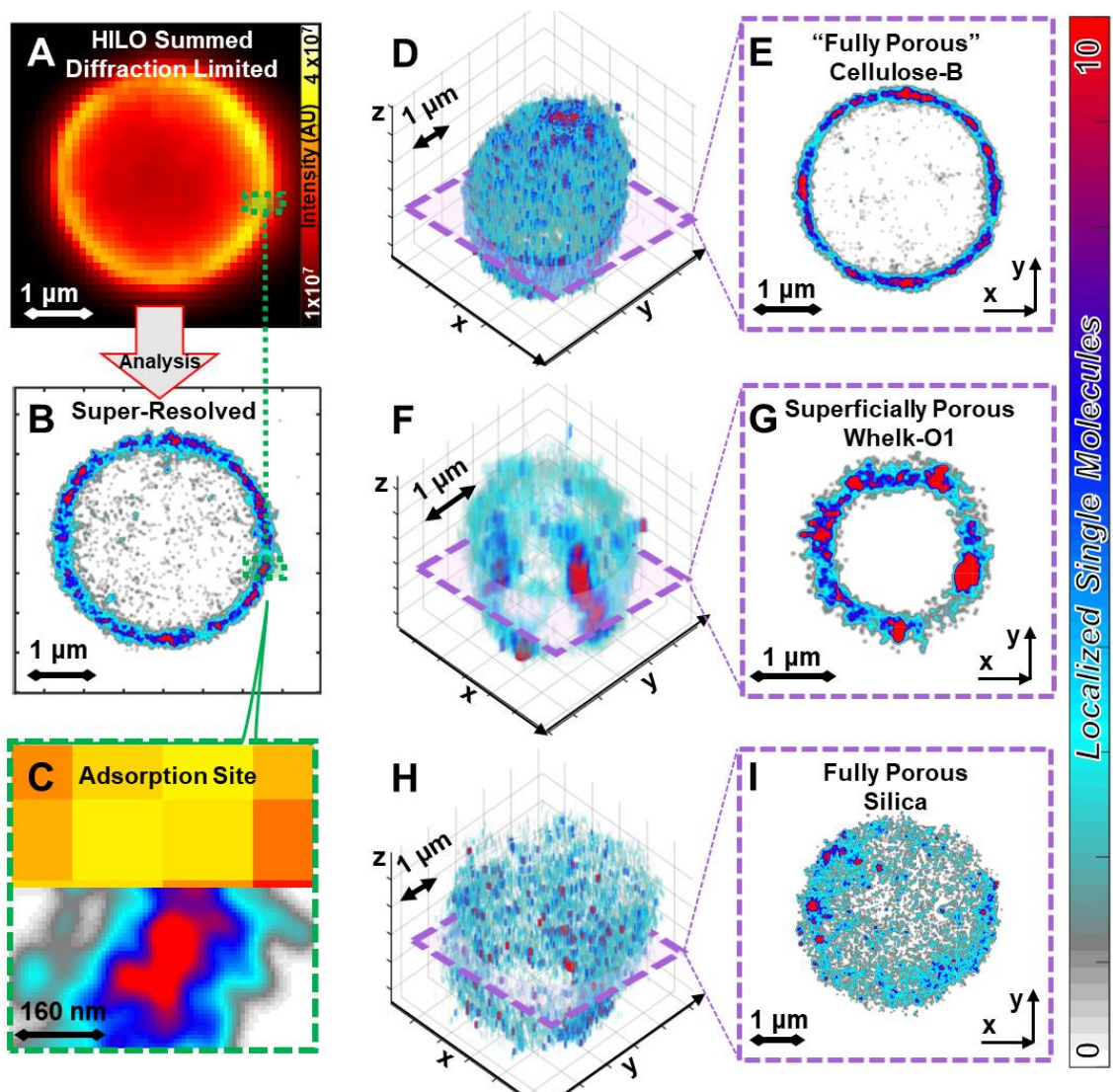

**Fig 2. Super-resolution imaging reveals resistance to mass transfer in porous Cellulose-B chromatography particles.** (**A**) Diffraction-limited image of fluorescent analytes adsorbed onto chromatographic Cellulose-B stationary phases using a light-sheet (HILO; fig. S1) (**B**) Imaging is improved to 30 nm super-resolutions with single-molecule localization analysis, (**C**) allowing for characterization of the number of molecules at individual adsorption sites. (**D-I**) 3D maps of single molecule adsorption events collected in 250 nm axial steps, with example 2D slices at the half diameter for (**D-E**) Cellulose-B particles, (**F-G**) Whelk-O1 superficially porous particles, and (**H-I**) fully porous non-functionalized silica particles. See Fig. 1A and table S1 for more details. Access of analytes in the inner volume of the "fully porous" Cellulose-B is comparable to the superficially porous particles, not the fully porous, unfunctionalized silica, showing the limited accessibility to the pore network. Quoted "fully porous" is used to indicate difference between the marketed and our observed porosity.

The marketed "fully porous" stationary phase materials show a similar distribution of adsorption locations to those seen in the SPP design. Contrary to the commonly accepted image of diffusion through the porous chromatography particles, super-resolution imaging reveals that analytes cannot freely diffuse throughout the volume of the tested Cellulose-B and Whelk-O1 FPPs (Figs. 2D, 2E, and fig. S2). The super-resolution map of Cellulose-B is markedly similar to the Whelk-O1 SPP with a solid core (Fig. 2G) and does not resemble the silica FPPs (Fig. 2I). All the stationary phases show nanoscale heterogeneity in adsorption, with discernable high affinity and non-specific adsorption sites (red, blue spots in Fig. 2, respectively).

The similarity between the Cellulose-B and SPPs is quantified by taking the ratio of the single-molecule localization distance to the radial distance from the stationary phase particle edge, *r*, as a percentage (*%r*) (Fig. 3A). Qualitatively, the Cellulose-B and Whelk-O1 distributions show a sharp drop and sparse spatial distribution of analytes near the center compared to the silica FPPs. Quantitatively, we examine the *%r* that analytes are able to access and define an extreme limit of accessible distance into the particles, where 99% of analytes cannot reach. In

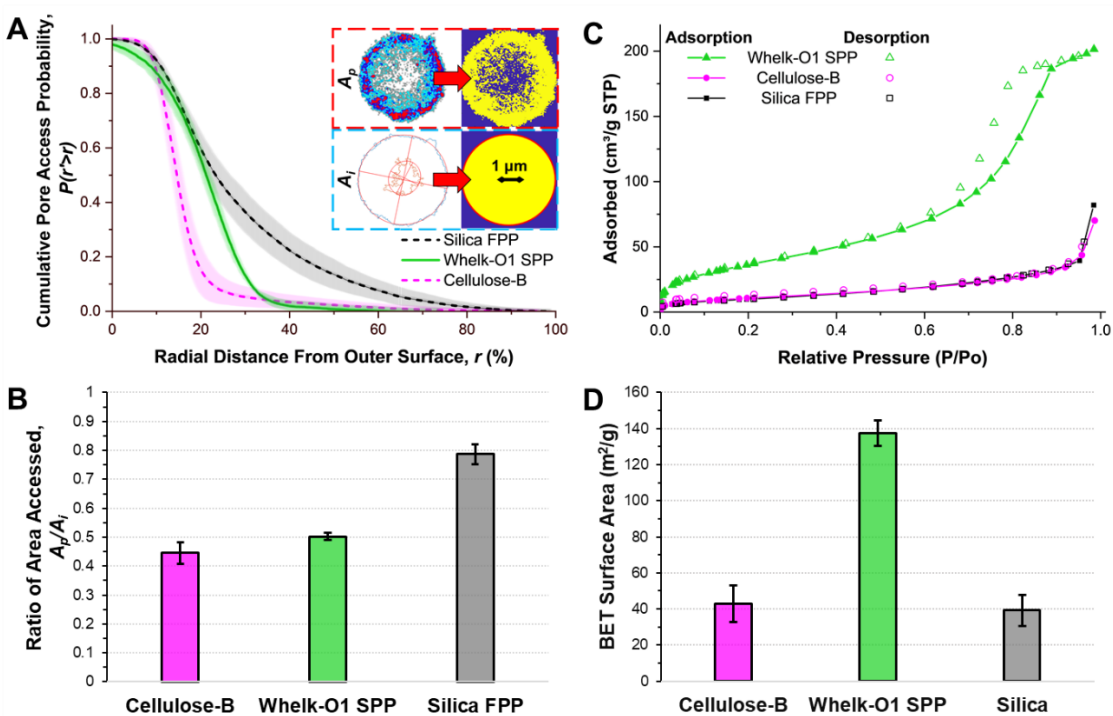

**Fig. 3. Single-molecule locations reveal spatial analyte pore accessibility otherwise hidden in *ex situ* and ensemble methods.** (**A**) The cumulative distribution of all identified single molecules as the distance from the outer edge of each particle, defined as a % of the particle radius *r*. Data are average (line) and standard deviation (shaded) from three different particles. Data on semi-log scale shown in fig. S3A. (**A**, inset) The relative accessible surface area from super-resolution imaging is calculated as the ratio of the area probed by analytes ($A_p$) over the imaged cross-sectional area ($A_i$), and plotted in (**B**). See fig. S4 for details. (**C**) Nitrogen adsorption-desorption isotherms obtained at -196 °C. (**D**) The surface area was extracted from $N_2$ isotherms using the BET method. The changes in accessible surface area for analyte adsorption revealed by *in situ* fluorescence in **B** are not resolved by the *ex situ* isotherms in **D**.

silica FPPs up to 84 ± 3*%r* is accessible to 99% of analytes (Fig. 3A, black). For Whelk-O1

SPPs, only 48 ± 1 %r is reached (Fig. 3A, green), matching the expected limit of ~ 37 %r set by the ratio of the 1.7 μm/0.5 μm core/shell design. Finally, for Cellulose-B only 64 ± 2 %r (Fig. 3A, magenta) is accessible, which if compared to the Whelk-O1 SPP cutoff, can equate to the Cellulose-B particles having a ~1.8 μm wide solid core and ~1.6 μm porous layer particle design. Notably, even Whelk-O1 FPPs showed similar spatial distributions to the Cellulose-B and Whelk-O1 SPPs (fig. S2D).

**The unexpected superficial porosity revealed by super-resolution imaging is obscured in traditional *ex situ* characterization methods**

Our single-molecule spatial distributions quantify the limited pore accessibility under solvated conditions that is otherwise not discernable by common characterization techniques. First, we translate our super-resolution information to values comparable to conventional methods by quantifying the cross-sectional area probed by analytes ($A_p$) per total imaged area of the stationary phase ($A_i$) (Fig. 3A, inset; fig. S4). We observe a clear discrepancy between the functionalized and non-functionalized particles; analytes only access 45 ± 4% and 50 ± 1% of the Cellulose-B and Whelk-O1 area, respectively, compared to 79 ± 3% for silica (Fig. 3B). Next, we perform gas adsorption isotherms combined with BET (Brunauer, Emmett, Teller) analysis (Fig. 3C, supporting information s9), one of the most ubiquitous methods used to characterize porosity and surface area of chromatography materials (*7*). The ensemble-averaged isotherm results do not resolve any difference in the FPPs. Cellulose-B and silica have identical surface areas of 40 ± 10 m$^2$/g, matching the value reported as a specification by the manufacturer of the silica particles (*23*). However, the shapes of the isotherm curves are indicative of a non-adsorbing material, as is further elaborated on in the discussion. In contrast, the micro-porous Whelk-O1 SPPs have a characteristic "S" shape isotherm that corresponds to monolayer adsorption, resulting in 138 ± 7 m$^2$/g surface area (*24*). The BET results contradict the known particle design where the FPPs should have higher surface area than the SPP design. Finally, scanning electron microscopy (SEM) imaging of focused ion beam cross-sectioned Cellulose-B particles (supporting information s10) show no indication of pore blocking, with ~100 nm macropores clearly discernable on the surface and within the particles (fig. S7). Yet, our single-molecule measurements show that the large inner pores are inaccessible to our small ~1 nm analyte in functionalized FPPs on the time scale of a typical chromatography experiment (~mins-hrs).

**Removing functionalized cellulose coating increases pore accessibility**

Why are the "fully porous" Cellulose-B particles showing superficially-porous behavior with the rhodamine 6G analyte? We first ruled out that the low pressure of ~700 psi compatible with our microscopy setup as a cause since we observe a similar superficially porous spatial distribution in Cellulose-B at both increased and decreased pressures (fig. S9). Similarly, we used refractive index matching media to confirm that the functionalized material is not optically obscuring analytes within the particles (fig. S11). We then removed the cellulose polysaccharide coating by treating the particles with 10, 20, and 100% (v/v) dimethyl sulfoxide (DMSO) in H$_2$O (supporting information s14), which is a strong solvent for cellulose (*25*).

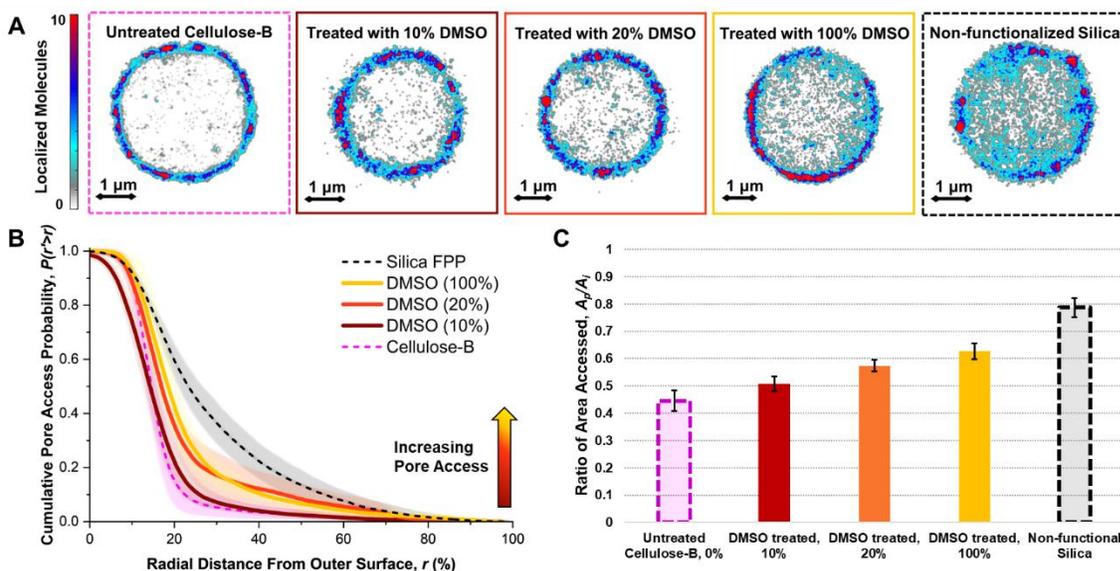

**Figure 4. Pore network accessibility in Cellulose-B particles is increased by treatment with organic solvent.** (**A**) Super-resolution maps show increased access to the inner volume of the particles after treatment in DMSO solvent at varying concentrations. Cellulose-B particles were treated over 24 hours at 40 °C. (**B**) The radial distribution quantifies the degree of recovery in pore access in the inner volume of the particles, approaching the original full pore volume of the silica gel backing as DMSO concentration is increased. Data are average (line) and standard deviation (shaded) from three different particles each. Data on semi-log scale are shown in fig. S3B. Lack of full recovery shows that some loss in pore volume will occur with any coating present. (**C**) Accessible area by the analyte increases consistently with stronger DMSO treatment, (i.e. decreasing degree of functionalization), showing a quantifiable measure of porosity in the same format as Fig. 3B.

Removal of the functionalized cellulose with increasing solvent strength increases pore accessibility in Cellulose-B (Fig. 4, fig. S12). Compared to the untreated Cellulose-B (Fig. 4A, magenta), samples treated with 10% DMSO (dark red) show a visible increase in analyte accessibility to the inner volume of the particles, with further improvements with 20% (orange) and 100% (yellow) solvent. Quantifying the spatial localizations, analyte accessibility inside particles increases with increasing solvent concentration (Fig. 4B). Notably, we successfully regain analyte access to the inside of Cellulose-B particles, reaching distances up to $86 \pm 3\%r$ (Fig. 4B and fig. S3B, yellow) with 100% DMSO treatment, statistically identical to the $84 \pm 3\%r$ observed for silica (Fig. 3A). Furthermore, we observe a consistent increase in available surface area with increasing solvent strength, reaching up to $63 \pm 3\%$ in 100% DMSO, which remains less than the $79 \pm 3\%$ in silica (Fig. 4C).

Time of Flight-Secondary Ion Mass Spectrometry (ToF-SIMS) imaging corroborates our super-resolution microscopy to determine that the functionalized polysaccharide coating blocks the pores in Cellulose-B. Elemental mapping of cross-sectioned particles reveal that the dense cellulose and chlorine-containing tris(3,5 dimethyl phenyl carbamate) groups are reduced primarily near the outer edges of the particles (fig. S13A). After treatment with 100% DMSO,

ToF-SIMS show the functionalization is only partially removed, leaving chlorine-containing groups throughout the particles (fig. S13B). We attribute the primary reason behind blocked pore accessibility in the Cellulose-B particles to a "shell" of functional groups forming near the outer surface of the particles, preventing analyte access to the inner volume.

**Analyte adsorption kinetics which underlie separation elution time are maintained with the removal of the cellulose coating**

We model and experimentally measure chromatographic separation to test the effects of removing some, but not all, of the functional groups with DMSO. The temporal data in our single-molecule measurements allows us to predict ensemble elution behavior. Molecular adsorption within the porous particles affects the larger scale separation process, where mass transfer and elution are dependent on the strength and duration of adsorption time. Our super-resolution maps (Fig. 2) include temporal information (Fig. 5A, inset) of the time individual molecules stay adsorbed on a site ($t_D$; dwell time) that are related to desorption rates ($k_d = 1/t_D$), along with association data. Our kinetic data aligns with the stochastic theory of chromatography (*26*), where individual adsorption sites have first-order kinetics, but heterogeneity between different of sites lead to an "n-site" distribution (*27*). Experimentally we measure adsorption free energies varying from - 2.9-11.4 kJ/mol, reasonable values for non-specific adsorption of the rhodamine 6G analyte to the stationary phase surfaces (fig. S14) (*28*, *29*). Using the stochastic theory with the Lévy process representation (supporting information s18) and all adsorption events within the particles (Fig. 5A), we model the effects of the heterogeneous kinetics on the anticipated ensemble elution profiles (Fig. 5B, top) (*3*, *30*).

Single-molecule adsorption kinetics show that DMSO treatment does not change peak shape significantly while decreasing elution time (Fig. 5B, top). The single-molecule dwell times do not shift significantly after treating Cellulose-B particles with DMSO solvent (Fig. 5A, yellow), as the cumulative distribution curve shape remains close to that of untreated particles (Fig. 5A, magenta). This kinetic behavior further supports that functional groups remain even after DMSO treatment. There is a reduction in rare long-lasting adsorptions, which are associated with elution peak tailing and slower elution in a chromatography column. Our modeled curves reflect this change where the elution occurs earlier for DMSO treated particles (Fig. 5B, top).

Larger-scale, bulk HPLC separations using commercial packed chromatography columns validate our modeled elution curves. The HPLC elution peaks show strikingly similar trends to our model for a Cellulose-B column before and after treatment with DMSO solvent (Fig. 5B, fig. S16), where elution time is faster and tailing due to strong adsorption (or resistance to mass transfer) is not significantly changed. While the relative degree of the temporal shift in bulk does not exactly match the predicted changes from the model elution, our single-molecule experiments successfully predicted the faster elution and general peak tailing shape (fig. S16) observed in the bulk separation only using dwell times. The band broadening observed in HPLC includes additional effects of column packing and Eddy and longitudinal diffusion.

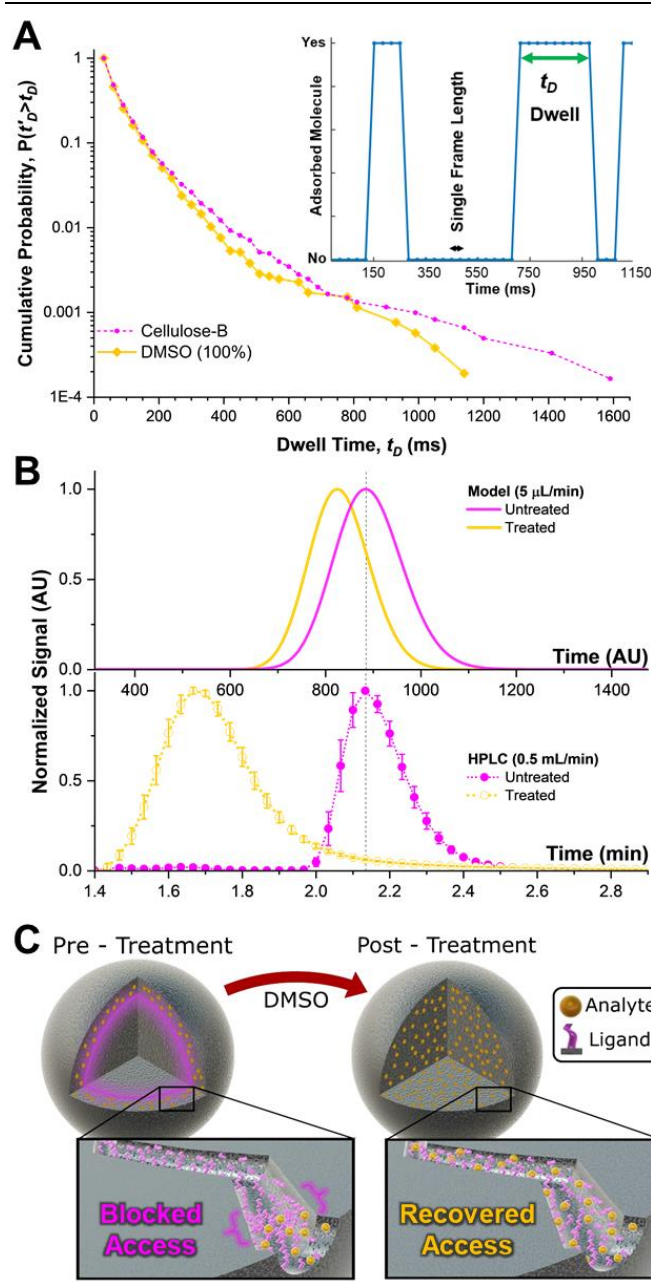

**Fig. 5. Single-molecule adsorption kinetics in porous stationary phase particles can reveal differences in scaled chromatographic elution.** (**A**, inset) Single-molecule kinetics of adsorption are measured as the time analyte molecules spend stuck (dwell time, $t_D$) at individual nano-scale adsorption sites throughout the imaged chromatography particles. (**A**) Cumulative distributions of measured dwell times for thousands of single-molecule adsorptions observed over at least three chromatography particles of each type. (**B**, top) Removal of the outer functional layer of Cellulose-B particles with DMSO treatment reduces overall retention time, while maintaining longer specific adsorptions not present in bare silica. (**B**, bottom) Experimental elution peaks of rhodamine 6G analyte, measured for commercial chromatography columns (REFLECT C-Cellulose B, Regis) on a HPLC system under comparable solvent and analyte conditions as in single-molecule measurements. Both the model and experimental HPLC elution peaks show tailing characteristic of strong adsorptions, with faster elution (shift to the left) after DMSO treatment of Cellulose-B. (**C**) Cartoon representation of mechanism where dense functional groups block access to the inner volume of the porous functionalized particle, which are then removed by DMSO treatment.

## Discussion

"Fully porous" stationary phase materials are over-functionalized leading to a reduction in analyte-accessible surface area that can be recovered with the removal of excess ligands (Fig. 5C). The addition of the adsorptive ligand materials to FPPs – either the polysaccharide for Cellulose-B or small-molecules for the Whelk-O1 FPP – reduces the surface area available for analyte mass transfer. The Cellulose-B and Whelk-O1 FPP data closely resemble that of the Whelk-O1 SPPs, not the FPP base silica (Figs. 2, 3, S2). This behavior surprisingly does not match the marketed "fully porous" material description. Elemental ToF-SIMS data (fig. S13) supports that excessive ligands block pores as a "shell" so that the interior of the particles is inaccessible to analytes (Fig. 5C, left), effectively forming an inefficient core/shell of 1.8 μm/1.6 μm size for Cellulose-B.

We importantly show a path forward for future stationary phase design to improve mass transfer. We clearly identify that the well-known challenges in ligand utilization and accessibility for resin-based purification of biologic analytes (31–34) are present for our significantly smaller ~1 nm analyte. Removal of excessive functionalization by solvent treatment in Cellulose-B increases pore accessibility while maintaining the respective adsorption kinetics (Fig. 5C, right). Engineers and manufacturers that functionalize bare stationary phases could optimize the ligand loading or explore functionalization strategies that maintain pore access. Coupling improved functionalization strategies (35) with improved stationary phase morphology (monoliths or membrane adsorbers) (36) could further increase ligand accessibility; however, changes to stationary phase morphology may impact the surface area and capacity (34).

The change in accessibility and SPP behavior of the Cellulose-B particles would never be apparent with diffraction-limited imaging. Past diffraction-limited scanning confocal microscopy produced static images that could not resolve adsorption site locations or any dynamics, only discerning larger protein analytes at the surface- vs. the interior of stationary phases (21, 22). Even our diffraction-limited light sheet image shows fluorescence intensity in the interior of the Cellulose-B, and hence incorrect qualitative access of the pores (Fig. 2A). Only when we achieve quantitative details by localizing dynamic analytes at 30 nm and 32 ms spatiotemporal resolutions do we determine the unexpected inaccessibility of over half the volume of the Cellulose-B. Further, visualizing the dynamic kinetics over a wide field of view importantly determines the adsorptive capabilities of the stationary phase before and after solvent treatment.

*Ex situ* methods that are the standards in stationary phase characterization cannot resolve the reduced analyte-accessible surface area in the functionalized FPP stationary phases. First, nitrogen adsorption isotherms with BET analysis remains widely used (37), but the theory relies on fundamental assumptions such as gas monolayer formation as the first adsorption layer (38). Inspection of the Cellulose-B and silica isotherms exhibit negligible adsorption, indicating the measured surface area is due to weak adsorbate-adsorbent interactions, rather than adsorption at surface sites. The sharp increase in adsorption at high partial pressures could be related to progressive multilayer formation from $N_2$-$N_2$ interactions, and so the assumption for BET analysis is not fulfilled (39). Therefore, the nitrogen adsorption isotherms incorrectly reveal identical surface areas for silica and Cellulose-B. Regardless, vendors still provide surface area information (23) obtained from an incorrect model. Furthermore, requiring dry samples can structurally change the hydrophilic stationary phases and the small size of the $N_2$ probe inherently overestimates the surface area accessible to larger analytes (40). SEM imaging confirms the presence of 100 nm wide pores within the Cellulose-B (fig. S7) but cannot discern the limited pore accessibility. SEM is carried out under vacuum where functional groups (and consequently sorption sites) are not identifiable, without any analytes of interest, and ignoring factors like pore swelling that occur in solution.

In contrast to *ex situ* techniques, *in situ* super-resolution fluorescence microscopy is non-destructive, can probe diverse analyte sizes, and is done in solvated conditions that can be tuned (pH, polarity, etc.) to match those used in the larger-scale separations run on columns. We note a limitation is the requirement of a photoluminescent signal from the probe, but fluorescent molecules can be strategically selected to mimic desired small molecule analytes or bioconjugated to larger biologics.

Imaging commercial materials demonstrates that the improved mass transfer observed by single-molecule methods (*41*) is directly translatable to real HPLC separations. Our results are the first agreement between single-molecule mass transfer and HPLC data. Prior single-molecule imaging of chromatography mainly used model surfaces mimicking stationary phases and solely reported single-molecule data (*6*). Only two groups have made direct comparisons between single-molecule and chromatography elution profiles, but required adding an unobservable component to the dwell time distributions (*42*) or construction of a simplified chromatography setup (*43*, *44*) to reach agreement. Here, using the exact same materials for single-molecule imaging and a commercially available column, we achieve remarkable overlap in peak shape and change in the trend in elution time between single-molecule and HPLC data (Fig. 5B, fig. S16). While the changes in peak broadening do not entirely match, this is to be expected as single-molecule measurements only probe adsorption and not the complicated diffusion that occurs within a packed column.

Super-resolution imaging of the spatial pore accessibility and temporal adsorption kinetics resolves the molecular phenomena underlying the reduced mass transfer term (*C*) of the commonly-used van Deemter equation (*45*): 1) the overall resistance to mass transfer that occurs throughout a column (influenced by porosity and surface area), and 2) the underlying slow adsorption-desorption processes (which are critical in areas like difficult chiral separations) (*46*–*48*). The van Deemter approach averages many behaviors into constants to determine chromatographic parameters by trial-and-error (*3*, *4*). The direct visualization and quantification of the molecular phenomena that occur within a column informs the design of stationary phases from the bottom-up. Specifically, here we show that reducing the degree of functionalizing stationary phase particles with cellulose and ligands can increase available porous surface area (influencing resistance to mass transfer), without disturbing the underlying adsorption kinetics (maintaining selectivity). Future work can develop methods to image the Eddy diffusion on a packed particle or membrane systems, as well as improve microscope spatiotemporal capabilities.

**Acknowledgments**

**Funding:**
The Case Western Reserve University College of Arts and Sciences and RCSA Cottrell Scholar fund provided financial support for this work.

**Author contributions:**
R. M. N.: Overall conceptualization, methodology, software, validation, formal analysis, investigation, data curation, writing-original draft, writing – review & editing, visualization, supervision, project administration.
M. Z.: Gas adsorption isotherm conceptualization, validation, formal analysis, investigation, data curation, writing – original draft, visualization.
A. K.: Flow rate experiments investigation and data curation, and pressure analysis conceptualization, visualization, and writing – original draft.
T. K. J. K.: ToF-SIMS elemental mapping experiments conceptualization, investigation, methodology, software, data curation, visualization, writing – review & editing.
N. H.: Ion beam milling for ToF-SIMS conceptualization, investigation, methodology, validation, writing – original draft.
J. S. P.: SEM imaging and FIB conceptualization, investigation, methodology, data curation, visualization, writing – review & editing.
B. G.: Gas adsorption isotherm conceptualization, supervision, resources, writing – original draft.
C. E. D.: Overall conceptualization, supervision, writing – review & editing.
R. A. S.: HPLC packed column experiments conceptualization, methodology, validation, formal analysis, investigation, data curation, resources, writing – original draft, and writing – review & editing.
L. K.: Overall conceptualization, methodology, validation, writing-original draft, writing – review & editing, supervision, project administration, resources, funding acquisition.

The authors thank Case School of Engineering's Swagelok Center for Surface Analysis of Materials at Case Western Reserve University for use of equipment, analysis, and sample preparation.

We also thank Andrew Lininger and Prof. Giueseppe Strangi for helpful discussion and assistance with figure preparation, Regis Technologies and Dr. Edward G. Franklin for discussion and providing the chiral stationary phase materials, and Lianna Johnson and the Kisley group for useful discussion. L.K. and R.A.S. thank Prof. Kristin Cline for establishing their initial collaborations in chromatography.

**Competing interests:**
The authors declare no competing interests.

**Data and materials availability:**
All data is available in the main text or supplementary text.

**Supplementary Materials**
Materials and Methods
Supplementary Text

References (49-59)
Supplementary Data (.xlsx)
Figs. S1 to S16
Tables S1 and S2

**Supporting Information**
**Super-resolution imaging reveals resistance to mass transfer in functionalized stationary phases**

Ricardo Monge Neria,[1] Muhammad Zeeshan,[2] Aman Kapoor,[1] Tae Kyong John Kim,[3] Nichole Hoven,[3] Jeffrey S. Pigott,[3] Burcu Gurkan,[2] Christine E. Duval,[2] Rachel A. Saylor,[5] Lydia Kisley[1,4*]

[1]Department of Physics, [2]Department of Chemical and Biomolecular Engineering, [3]Swagelok Center for Surface Analysis of Materials, and [4]Department of Chemistry, Case Western Reserve University, Cleveland, Ohio 44106-7079
[5]Department of Chemistry and Biochemistry, Oberlin College, Oberlin, OH 44074
*lydia.kisley@case.edu, ORCID: 0000-0002-8286-5264

**Table of Contents**

## s1. Commercial Chromatography Materials Used.

We studied stationary phase materials that vary in porosity and chemical functionalization as detailed in Table S1 to complement the details provided in Figure 1.

**Table S1. Physical properties of commercial stationary phase particles, as provided by the suppliers.**

| Stationary Phase | Base matrix | Pore diam. (nm) | Particle diam. (µm) | Functionalization | Vendor | Additional Detail |
|---|---|---|---|---|---|---|
| Cellulose-B | Fully-porous silica | 100 | 5 | Cellulose tris(3,5 dimethyl phenyl-carbamate) | Regis | … |
| Silica | Fully-porous silica | 100 | 5 | None | Glantreo | … |
| Whelk-O1 SPP | Superficially-porous silica | 10 | 2.7 | 1-(3,5-Dinitrobenzamido)-1,2,3,4,-tetrahydrophenanthrene | Regis | 1.7 µm solid core; 0.5 µm porous shell |
| Whelk-O1 FPP | Fully-porous silica | 10 | 3.5 | 1-(3,5-Dinitrobenzamido)-1,2,3,4,-tetrahydrophenanthrene | Regis | … |

## s2. Sample Preparation

Commercial stationary phase particles were immobilized on silica glass for microscopy by relying on polar interactions, as previously described.([13]) Microscopy slides (20 mm x 30 mm, #1.5, Fisherbrand) were prepared by submerging in a base-peroxide ($H_2O_2$ + $NH_4OH$ + H2O, at a 1:1:6 volume ratio, respectively) bath at 70 °C for 90 s, water rinsed, dried with nitrogen gas (5.0 grade, Airgas), then plasma cleaned in an $O_2$ (Industrial grade, Airgas) plasma cleaner (PDC-32G, 115 V, Harrick Plasma) at 140-280 Torr, medium power, for 2 minutes. The base-peroxide bath was prepared with 30% Certified ACS, Thermo Scientific, $H_2O_2$, and Certified ACS Plus, Fisher Chemical $NH_4OH$. All uses of water in this work utilized Type 1 ultrapure water purified on an Elga Chorus 2+ system. Different water suspensions of 0.1 wt % solids were used to deposit cellulose tris (3,5-dimethylphenylcarbamate)-functionalized fully porous 5.0 µm silica stationary phase particles (Regis), non-functionalized 5.0 µm silica gel particles (SOLAS, Glantreo), 3.5 µm Whelk-O1 fully porous particles (Regis), and 2.7 µm Whelk-O1 superficially porous particles (Regis). The specific commercial particle parameters are summarized in Table S1. An 8 µL volume of the stationary phase solutions was then dropcast on the slides, where the cleaned surfaces were hydrophilic, allowing for physisorption by polar interactions with the 0.1% solids suspended in water. The slides were then covered by silicon flow cells (Hybriwell, 13 mm diameter, 0.15 mm depth, Grace Biolabs) and rinsed multiple times with water, then left to rehydrate for at least 2 hours before imaging.

**s3. Single-Molecule HILO Microscopy.**

Single-molecule adsorption was characterized by highly inclined and laminated optical sheet (HILO) fluorescence imaging of nanomolar concentrations of fluorescent rhodamine 6G dye (99%, Fisher) under flow, as previously described (*13*). The microscope body consists of an Olympus IX-73 inverted microscope, with a CNI diode laser (532 nm, MGL-III-100 mW) used for excitation. The laser light is guided through an achromatic doublet lens (visible, 50.8 mm, 400 nm EFL, Newport) mounted on a single axis translation stage (462 series, SM25, Newport) that allowed us to control the incident angle of the laser at the sample, set to a 77 ° exit angle to achieve HILO. A $z$ (vertical) single axis translation nano-positioner (Physik Instrumente, PIFOC, P-725) with a 400 μm travel range is paired with the 100x magnification oil immersion objective (Olympus, 100x, NA 1.49, UAPON100XOTIRF) to simultaneously control the imaging and illumination planes, allowing for 3D imaging. Imaging was done with 1 nM rhodamine 6G solutions in 20 mM HEPES buffer (Biotang Inc, pH 7.33), passed through the flow cells at a rate of 5 μL/min using a syringe pump (NE-1000, New Era Pump Systems Inc.). An EMMCD camera (iXon Life 897, Andor) collected the emission signal at 30 ms exposure, 400 EM gain, in Frame Transfer mode, with 532 nm laser excitation of 10 mW (379 W/cm$^2$) at the sample.

**s4. Single-Molecule Data Analysis**

Single-molecules were identified and analyzed using a home-written, publicly available (*49*) MATLAB (2022b) code that uses 2D Gaussian fitting strategies, paired with radial symmetry (*50*) centroid position refinements, to obtain super-resolved, single-molecule locations. We use the single axis nano-positioner to vary the vertical position of the imaging plane, first mapping the porous materials in 2D at each height, then stitching the planar slices as 3D voxels with a modified version of Woodford's *vol3d_v2* algorithm (*51*)**.** Further details can be found in Monge Neria and Kisley (*13*).

**s5. (fig. S1) Improving HILO Illumination.**

Signal to background is improved compared to our previous work (*13*) by the addition of an optical iris as a light stop on the illumination path, reducing the width and thickness of the light sheet. Similar to the characterizations detailed by Tokunaga et. al. (*52*), we measure the relative intensity ({signal – background} / {signal + background}) of fluorescent beads (Orange FluoSpheres, 0.2 µm diameter, 540/560 nm EX/EM, Invitrogen) immobilized by mixing within a 3% wt agarose (Sigma-Aldrich) hydrogel in water. The fluorescent beads were imaged under the same conditions as the single-molecule measurements of rhodamine 6G, with the 77 ° illumination exit angle to achieve HILO. An iris (Thorlabs, ID50/M 2.5 to 50.0 mm standard mounted iris diaphragm) placed right before the doublet lens was manually opened and closed to control the width of the illumination light sheet (figs. S1B-D), for which we achieve a minimum of 4.96 µm (fig. S1A). We record the average fluorescent intensity of the beads and background by scanning up in the z axis (-1 to 40 µm from the glass/gel interface) using the nano-positioner, then apply our single-molecule analysis algorithm. We define the width of the light sheet (*R*) by the full-width half-max of an illumination profile (fig. S1E), obtained by taking a linear cross-section of the intensity for the summed raw data over a full scan (fig. S1C) at different iris apertures. Note that field divergence causes higher intensity near the edges at larger iris apertures. Linear extrapolation of the sheet width vs intensity profile was used to estimate the 116.5 µm max sheet width, based on the other measured profiles. From plotting the analyzed single molecule relative intensity data, we estimate the thickness of the light sheet (*dz*) as the estimated full width half-max of the data near the interface (fig. S1F). However, we do note that the use of a circular light-stop (rather than slits), and the use of a single area scan, does contribute to the noisier background signal and larger shifts in illumination seen at larger *R* widths (fig. S1 F, insets). Finally, we plot out the measured light sheet thickness (*dz*) with respect to the measured width (*R*) of the profile and compare to the theoretical relation proposed by Tokunaga et. al. ($dz = R/\tan(\theta)$; $\theta = 77$ °) (fig. S2G). From this we see that our light sheet matches the theory within error. We achieve a 2.3 ± 0.5 µm light sheet thickness for the operating ~11.36 µm width used for data collection when imaging stationary phase particles.

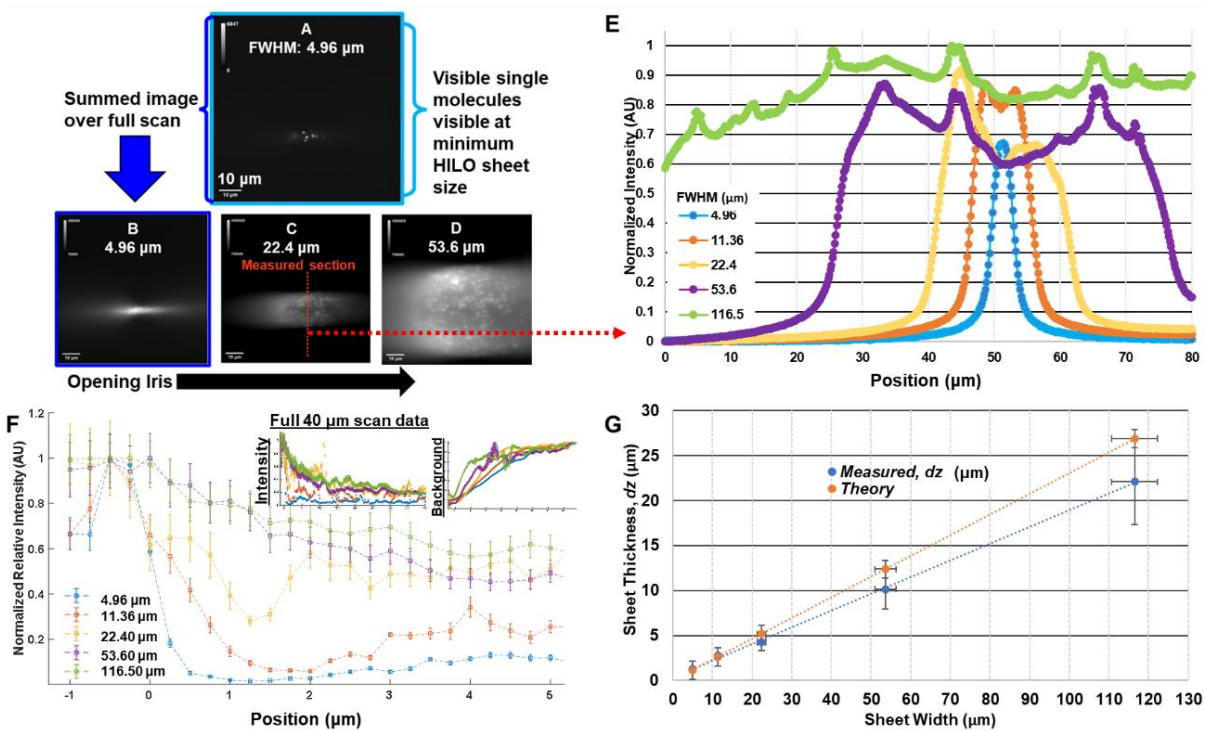

**fig. S1. Imaging immobilized beads in agarose gel for HILO illumination characterization.** (**A**) Fluorescent beads immobilized in agarose gel, imaged near the glass/gel interface at the minimum iris aperture, which results in a light sheet width (illumination area) 4.96 µm across. (**B-D**) Summed images of bead intensity raw data over a full 41 µm vertical scan through the gel sample, at varying iris aperture settings (i.e. varying sheet widths). (**E**) Width of illumination (*R*) was estimated by the width of the cross-sectional profile of the summed images, as shown in **C**. (**F**) Sheet thickness (*dz*) was approximated by the full width, half-max (FWHM) of the bead relative intensity profiles near the interface. (**F**, insets) Zoomed out full scan relative intensity profile, and the increasing background away from the interface. (**G**) Our measured sheet thickness (*dz*) vs width (*R*) compared to the theoretical *dz = R/tan(θ)*, where θ = 77°. Note that the FWHM for incomplete profiles was estimated by doubling the half-width measured.

**s6. (fig. S2) Single-molecule results for Whelk-O1 FPPs**

Just like the rest of the data displayed in the main text, we collected single-molecule scans of fully porous Whelk-O1 (Regis) particles using our rhodamine 6G probes. The 3D super-resolution mapping (fig. S2A) shows similar spherical structure to all other imaged particles, but a lower overall number of adsorbed molecules, which visually results in a more grayish-cyan map, with very few distinguishably strong adsorption sites (red). We can partially attribute this to the smaller particle size (3.5 μm), and the possibility of some size-exclusion behavior occurring with the nominally 10 nm wide pores. Notably, when looking at a 2D slice (fig. S2B) we observe a very similar spatial distribution of localized molecules to those observed in Cellulose-B fully porous particles (Fig. 2), and quantitatively similar accessibility to the Whelk-O1 SPP and Cellulose-B particles (fig. S2C). However, upon further quantification of these distributions (fig. S2D), it becomes evident that this behavior more closely matches that of Cellulose-B particles treated with 100% DMSO solution. This suggests that this loss in pore accessibility could be attributed to pore filling caused by functionalization, instead of primarily a dense "shell" of functional groups near the particle surface (as we've seen with the Cellulose-B particles). The Whelk-O1 FPPs have smaller pores that are chemically functionalized with a smaller ligand than the large polysaccharide. Therefore, these results show any type of functionalization can result in loss of porosity and can strikingly lead to superficially porous behavior in "fully porous" particles.

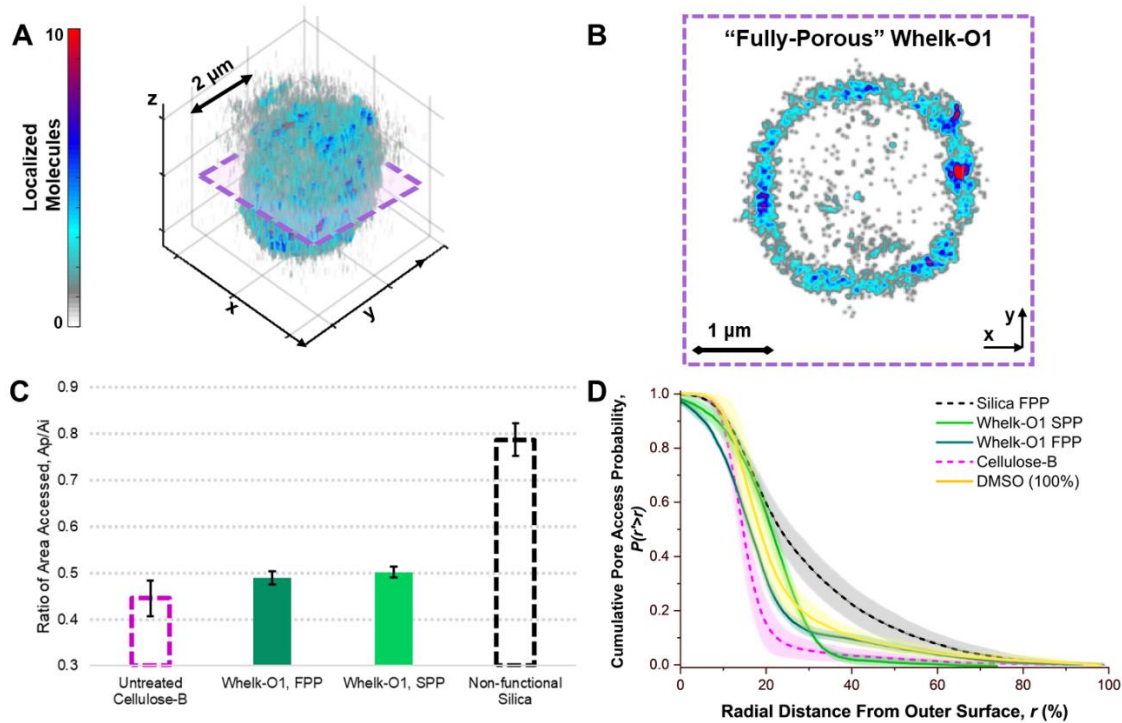

**fig. S2. Single-molecule fluorescence characterization of Whelk-O1 FPP particles shows pore blocking behavior comparable to a Cellulose-B FPP.** (**A**) 3D super resolution map of single molecule adsorption events on a 3.5 μm Whelk-O1 fully porous particle (FPP), generated from 2D slices, scanning up in 250 nm steps. (**B**) 2D slice of single molecule adsorption at the approximate half-height of each particle. Access of analyte at the inner volume of the fully porous stationary phase is comparable to the Whelk-O1 SPP and Cellulose-B (Fig. 2), showing limited accessibility to the porous network. (**C**) The relative accessible area is estimated as the ratio of the area probed by single molecules ($A_p$) over the imaged cross-sectional area ($A_i$), showing very similar results to Whelk-O1 SPPs and Cellulose-B (Fig. 3). (**D**) The distribution of analyte molecule adsorption radially from the edge of imaged particles shows very similar "core-shell" cut-off between all functionalized particles, but a gradual decrease on fully porous non-functional silica. The overlap between Whelk-O1 FPPs and Cellulose-B particles treated with 100% DMSO solution support our conclusion that an outer shell of functional groups was blocking the Cellulose-B particles, but the presence of any degree of functionalization will result in loss of porosity.

**s7. (fig. S3) Experimental Data in Semi-Log Scale**

Alternative presentation of the single-molecule, cumulative spatial distributions as distance from the edge of measured chromatography particles presented in Figs. 2-4 in the main text. The distributions are presented with a log10 y-axis scale for better readability of the 0.01 (99% of analytes) axis cut-off discussed in the main text. The other point of distinction here is that the differences between the hard cut-off on the Whelk-O1 SPPs and the other FPPs is much more apparent, as seen by the sharp drop on the green curve in fig. S3A at ~61%r. The tail end of the averaged cumulative distributions can increase in variance, and drop sharply due to the rarity (i.e. sparse data points) of molecules reaching deep into a particle volume. Results for DMSO treated particles are also included here in fig. S3B for comparison.

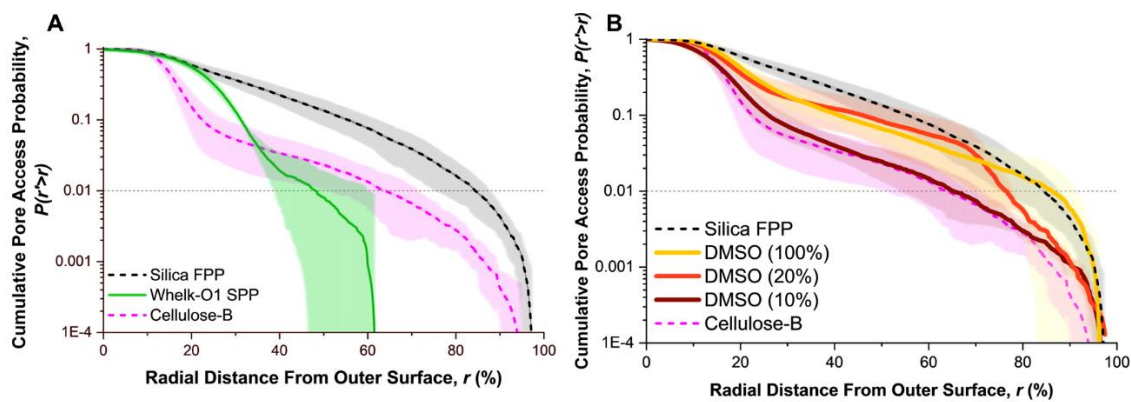

**fig. S3. Single-molecule fluorescence characterization of chromatography particles shows pore blocking behavior.** The log-linear cumulative distribution of all identified single molecules as distance from the approximated circumference of each particle, as a percent of each particle's estimated radius. Each line is generated from the average of 3 different imaged particles, with shaded regions corresponding to standard deviation. Semi-log plots of (**A**) data presented in Fig. 3, and (**B**) Fig. 4 in the main text.

**s8. (fig. S4) Stationary Phase Imaged Area vs Probed Area Approximations**

In order to quantify the approximate area of each particle that was successfully accessed by our rhodamine 6G analyte, we estimate the total particle surface area in view based on the super-resolution images. First, a super-resolution map (fig. S4A) is generated by plotting every single-molecule localization as a simulated 2D Gaussian, with a width defined as the ~25 nm estimated specific molecule position-uncertainty. The super-resolution map is then binarized using Matlab's *imbinarize* function (fig. S4B), and the area probed by single molecules ($A_p$) is estimated using the *bwarea* function. A contour plot of the binarized map is generated using the *imcontour* function (fig. S4C). The contour data is extracted, then fitted to ellipses using an edited version of Gal's *fit_ellipse* function.(*53*) Finally, for consistency, the cross-sectional area imaged ($A_i$) is estimated as the area of the fitted ellipse. The ratio of $A_p/A_i$ is then used to generate the data used for Figures 3B, 4C, S2C.

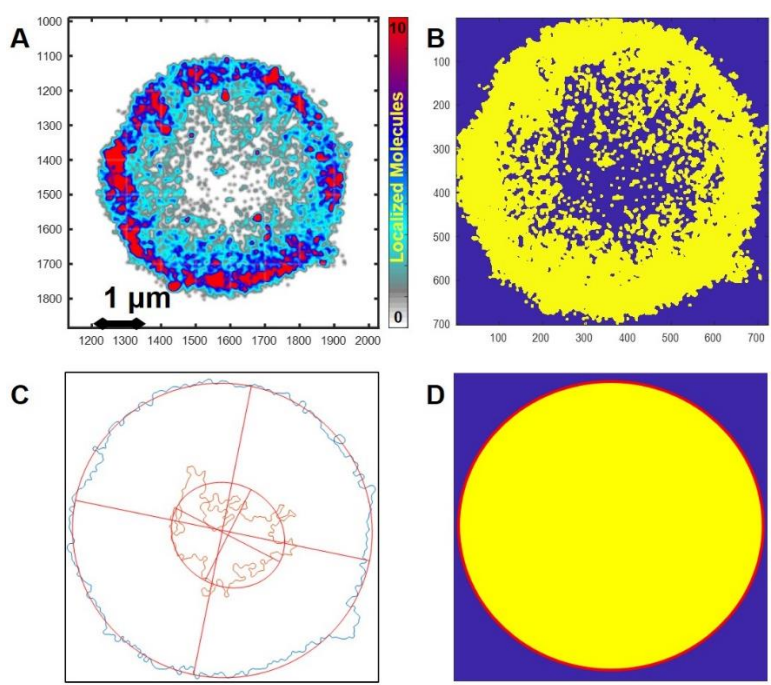

**fig. S4. The total particle area imaged, $A_i$, and the approximate area probed by the R6G molecules, $A_p$, are calculated by generating custom super-resolution maps and fitting ellipses to their binarized contours.** (**A**) The super-resolution map of a single imaged particle is generated from all the localized single rhodamine 6G probe molecules observed over 1000+ frames of collection. (**B**) The super-resolution map is binarized, from which the area probed ($A_p$) is estimated, then a contour plot is created from the binary image (**C**). The data from the contour is extracted, then fitted to ellipses using Gal's *fit_ellipse* function.(*53*) (**D**) The area of the larger outer ellipse is used as the estimate of $A_i$, the area imaged. Axis labels are in units of digital pixels, with each pixel corresponding to 8 nm in the physical sample.

## s9. (figs. S5 and S6) Nitrogen Adsorption Isotherms

Nitrogen adsorption isotherms for silica FPPs, Whelk-O1 SPPs, and Cellulose-B chromatography particles were collected. To examine thermal stability of the samples and ensure the absence of volatiles prior to nitrogen adsorption measurements, thermogravimetric analysis (TGA) was performed (TA instruments, Discovery TGA 55) by employing a ramp rate of 2 °C/minute from 25 °C to 500 °C under inert $N_2$ (5.0 grade, Airgas) atmosphere. Approximately 10 mg of a sample was placed in a platinum pan for each experiment. The obtained TGA curves of the samples are presented in fig. S5.

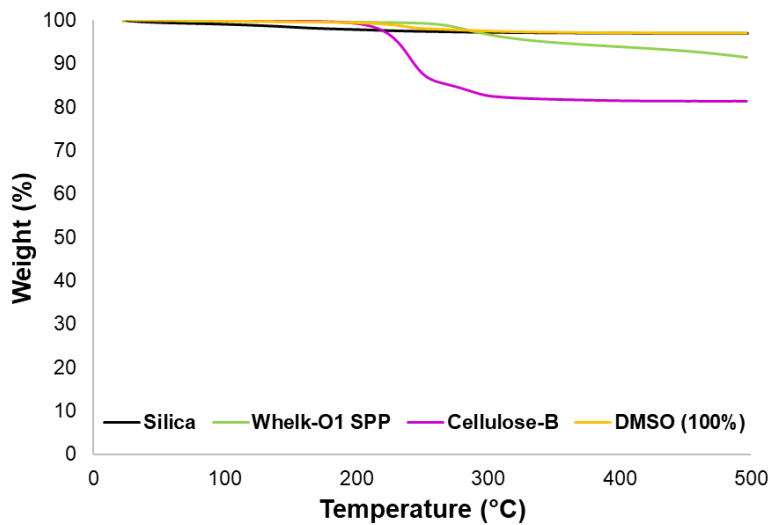

**fig. S5. TGA curve of the sample under $N_2$ atmosphere up to 500 °C, demonstrating good thermal stabilities.**

A Micromeritics TriStar II 3020 physisorption analyzer was used to determine the surface area of the studied samples through the Brunauer–Emmet–Teller (BET) analysis. Prior to each measurement, the samples were activated overnight at 150 °C under continuous flow of $N_2$ to ensure the removal of volatile residuals from the pores. Following the activation, $N_2$ adsorption–desorption isotherms were measured at -196 °C and the relative pressure range ($P/P_o$) of $10^{-6} - 1$ bar. For each analysis, approximately 150 mg of sample was used. The isotherm data in the partial pressure range of 0.05 to 0.3 were used in the BET Eq. (S1) ($7$) to estimate the surface areas of samples. Results (Fig. 3C) were reproducible within an experimental error range of ±10%.

$$V/V_m = C(P/P_o) \ / \ [1 \ + \ (C-1) \ (P/P_o)[1 - P/P_o)]$$  Eq. (S1)

$V/V_m$ represents the ratio of the volume of gas adsorbed ($V$) to the monolayer capacity ($V_m$), which is the volume of gas required to form a complete monolayer on the materials surface. $C$ is the BET constant, which is related to the energy of adsorption in the first adsorbed layer and indicates magnitude of the adsorbent-adsorbate interactions. $P/P_o$ is the relative pressure of the gas, where $P$ is the equilibrium pressure of the gas at a specific relative pressure, and $P_o$ is the saturation pressure of the gas. Data collection and analysis were carried out in duplicate, then averaged and error propagated to obtain the surface area measurements reported in Fig. 3D. We observed little to no difference in measured isotherm data between the Cellulose-B samples before and after treatment in 100% DMSO solution (fig. S6).

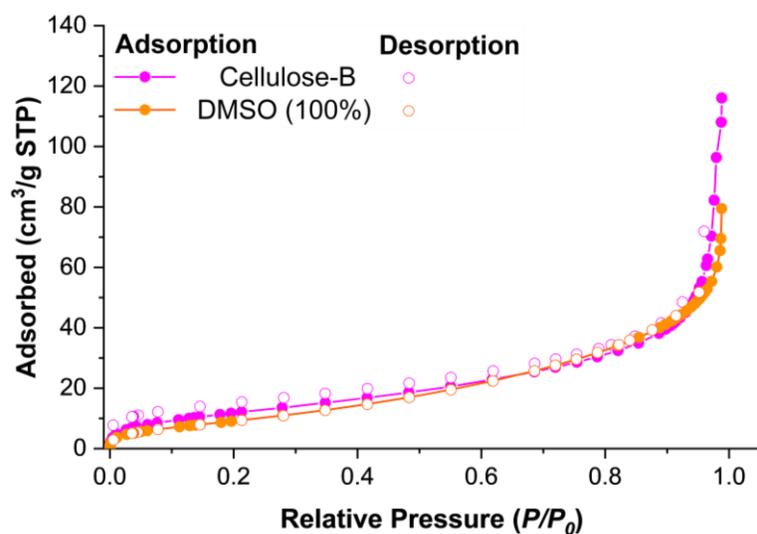

**fig. S6.** Nitrogen adsorption isotherms of Cellulose-B chromatography particles before and after treatment in 100% DMSO solution show little to no change in measured adsorption.

**s10. (figs. S7 and S8) SEM Particle Characterization and Size Distribution**

Scanning electron microscope (SEM) images of immobilized particles were collected using a Helios NanoLab 650 SEM. A small ~5 mg sample of dry particles was mounted on an aluminum SEM stub (1/2" slotted head, 1/8" pin, Ted Pella) using double-sided conductive copper tape (12.7 mm width, 3M™). The particles were deposited on the copper tape using a stainless-steel spatula, pressed lightly to ensure they adhered then the stub was tapped lightly at an angle to remove excess particles. The sample was prepared for SEM imaging and focused ion beam (FIB) milling by sputter coating a ~34 nm layer of Pd (Palladium Target, 99.95% Pd, Ø60 mm x 0.1 mm, from Ted Pella) using a Denton Vacuum Desk V turbo-pumped sputter coater. Secondary-electron images of single-layered particles on the tape surface (fig. S8) were acquired on a FEI Helios NanoLab 650 dual-beam SEM/FIB using an accelerating voltage of 10 kV and probe current of 0.2 nA (fig. S8 C).

A cross-section of a single particle (fig. S7A) was milled using the FIB on the Helios NanoLab 650. Before milling, a protective cap of Pt was deposited on top of the single particle to reduce surface damage from the ion beam. (see top of fig. S7 A). The Pt protective cap was deposited using an FEI Pt Deposition (CH3)3Pt(CpCH3) Gas Injection System. The Pt layer was deposited over an area of 7 μm (X) x 2 μm (Y), and 1.5 μm (Z) at ion beam conditions of 30 kV and 80 pA for ~9 minutes. Both milling and Pt deposition were enabled by the use of the focused Ga$^+$ ion beam. The first bulk milling was done using a regular cross-section at 30 kV and 0.79 nA for ~5 minutes over an area of 10 μm (X) x 10 μm (Y) x 5 μm (Z). The bulk milling was followed by 3 polishing steps using cleaning cross-sections at 30 kV and 80 pA for ~5 minutes each over an area of 10 μm (X) x 1.5 μm (Y) x 5 μm (Z). As shown with a linear section (fig. S7B), the nominal ~100 nm pore widths can be measured / observed from imaging of the particle interior.

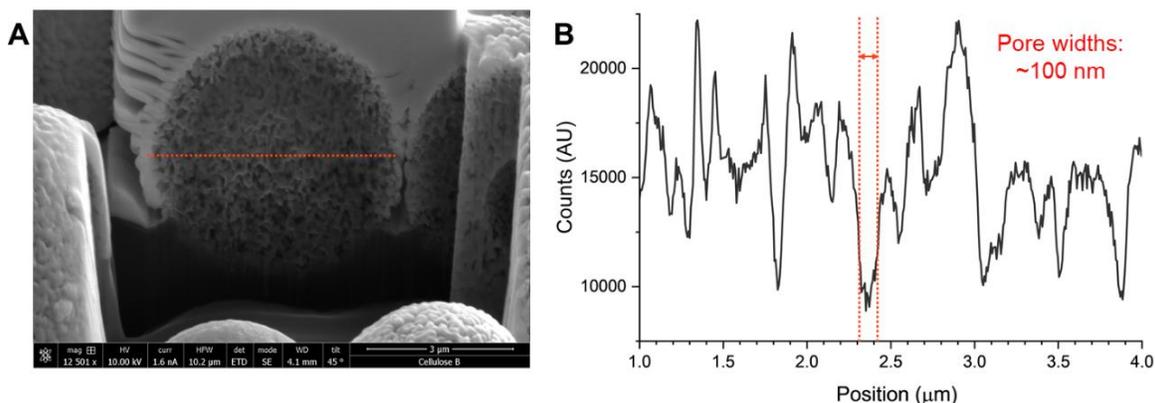

**fig. S7. Scanning electron microscope (SEM) image of a cross-section of Regis Cellulose-B commercial chromatography stationary phases shows fully 100 nm porous structure.** (**A**) A cross section of a single chromatography particle. Sectioning achieved by focused ion beam milling after depositing a Pt cap on top of the particle. (**B**) The pore structure is discernable and pore sizes measurable.

The high-resolution SEM images provide a good basis for measuring particle sizes, allowing us to generate a particle size distribution (fig. S8D). A SEM capture showing a large population of particles was selected (fig. S8C), then the size of each particle was approximated as an ellipse using ImageJ's analysis function 'Analyze Particles' after thresholding (*54*). From the obtained individual particle area measurements, the estimated diameters were simply calculated by approximating each particle area as a circle. The resulting distribution shows a mean particle diameter of 5.0 ± 0.8 μm, matching the vendor's reported nominal value of 5 μm.

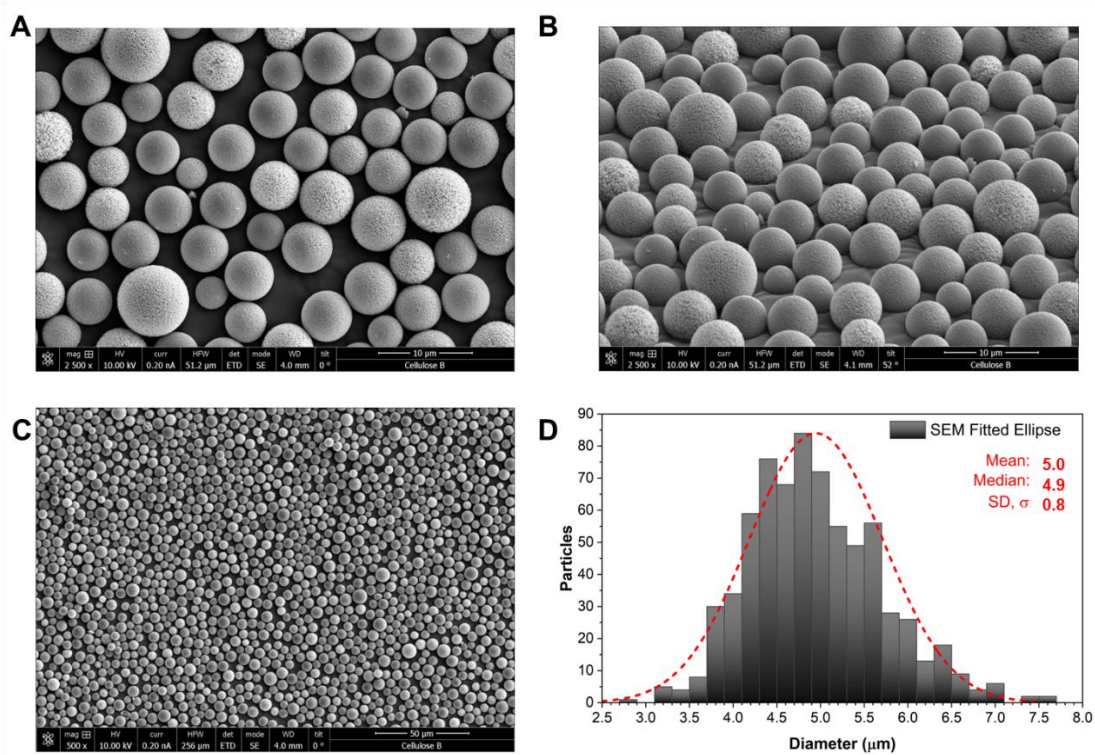

**fig. S8. Scanning electron microscope (SEM) images of Regis Cellulose-B commercial chromatography stationary phases. (A)** A top-down view of chromatography particles, where the surface pore structure is discernable and pore sizes measurable. (**B**) View at 52° tilt of the same particles shows further detail and morphology, and visualizes differences in particle sizes. (**C**). Image of a higher density of single particles allows for estimation of overall particle sizes. (**D**) The particle size distribution from SEM images of SPs was estimated using ImageJ's 'Analyze Particles' tool. A binary image is first generated from the SEM image and cleaned up using the drawing tools. Then, an image of the particle outlines is generated, and these are fitted to ellipses using the analysis tool, from which the particle diameters are calculated by approximating the measured ellipse area as the area of a circle.

**s11. (fig. S9) Flow Rate Does Not Change Analyte Accessibility**

We tested the effects of flow rate on single-molecule adsorption by carrying out the same overall experimental procedure on untreated Cellulose-B porous particles, flowing 1nM rhodamine 6G solution (in pH 7.33 HEPES) at various pump-driven flow rate settings and collecting 2000 frames at 32 Hz. The flow rates were varied from 0-100 µL/min which are roughly estimated to be at ~700 psi (using Eq. S4, see below), similar to conditions the Reflect columns are operated at, which can be as low as 400 PSI, and as high as 6000 PSI depending on mobile phase and application (*55*, *56*). We show that changing flow rate does not significantly affect the distribution of analytes within the volume of particles (fig. S9A), and in fact can decreases the overall analyte accessibility (fig. S9B). However, we note that the increasing the flow rate (expectedly) linearly increases the number of molecules that are observed over time (fig. S9C), since more volume and analytes are passed through during the same given time period. The fact that analytes do not more readily access the inner volume of the porous particles even at purely Brownian motion conditions (0 µL/min; i.e. no pump) suggests that the inaccessibility of the inner volume is not due to our chosen flow rate conditions.

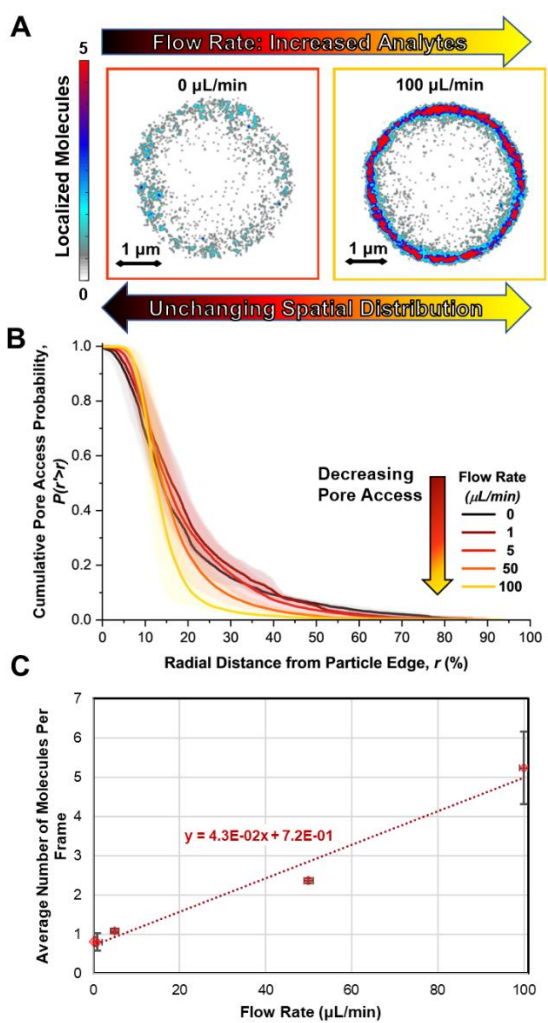

**fig. S9. Spatial distribution of analytes within the area of particles remains SPP-like with changing flow rate.** (**A**) The super-resolution map of imaged isolated porous Cellulose-B particles shows that more analytes are adsorbed on average, but the spatial distribution is unchanging, with the majority of analytes failing to enter the inner volume of particles. (**B**) The average number of analytes detected increases linearly with flow rate. (**C**) The cumulative distribution of analytes within the particles shows that accessibility to the inner volume decreases with increasing flow rate, but cannot be significantly increased, even under no directional flow. Flow rates 0, 1, 5, 50, and 100 µL/min correspond to approximately 677, 678, 681, 713, 750 psi pressures at the sample.

## s12. (fig. S10) Grace BioLabs Hybriwells Flowcell Pressure Calculation

To connect our flow-rate based mobile phase conditions inside our flow-cell system (Hybriwell, 13 mm diameter, 0.15 mm depth, Grace Biolabs) to chromatography relevant conditions, we derived a simple method for estimating the pressure inside the cell based on the flow rate set by our syringe pump (NE-1000, New Era Pump Systems Inc.), using 1/32" O.D. x 0.005" I.D. tubing (Tub Peek Red, IDEX Health), and 4.69 mm I.D. syringes (1mL, HSW – Norm Ject, Sigma).

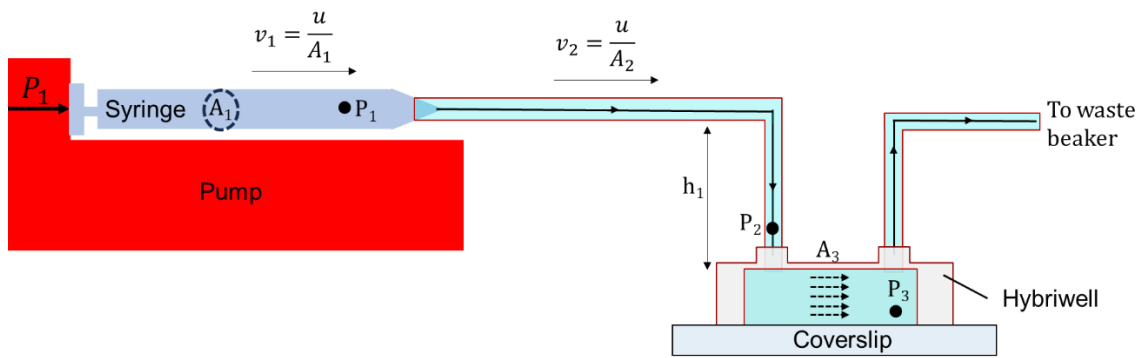

fig. S10. Illustration of the Grace BioLabs Hybriwells setup involving the syringe pump.

**Table S2. Variables used in pressure approximation.**

| Variable | Description | Value |
|---|---|---|
| $u$ | Volumetric flow rate of syringe pump | 1, 5, 50, 100 µL/min |
| $P_1$ | Pressure delivered by syringe pump (this is interpolated from data from the manufacturer) | 677, 678, 681, 713, 750 psi |
| $A_1$ | Cross-sectional area of syringe (known from syringe inner diameter) | $1.73 \times 10^{-5}$ m$^2$ |
| $v_1$ | Flow velocity in syringe | N/A |
| $h_1$ | Height drop from syringe pump to flow cell | 0.30 m |
| $P_2$ | Pressure in tubing just before entering flow cell | N/A |
| $A_2$ | Cross-sectional area of tubing (known from tubing inner diameter) | N/A |
| $v_2$ | Flow velocity in tubing | N/A |
| $P_3$ | Pressure inside hybriwells | 677, 678, 681, 714, 750 psi |
| $A_3$ | Cross-sectional area of hybriwells (known from manufacturer details) | $1.95 \times 10^{-6}$ m$^2$ |
| $v_3$ | Velocity inside hybriwells (we assume laminar flow here) | N/A |
| $\rho$ | Density of water | 1000 kg/m$^3$ |
| $g$ | Gravitational constant | 9.81 m/s$^2$ |

We use Bernoulli's Law to find the pressure in the tubing just before entering the flow cell based on the initial flow conditions from release in the syringe from a height $h_1$ above the flow cell: $P_1 + \frac{1}{2}\rho v_1^2 + \rho g h_1 = P_2$. Where is the density of solution, and g is the gravitational constant. To a first order approximation, we use continuity and the fact that $v = \frac{u}{A_3}$ to write:

$$P_2 = P_1 + \rho g h_1 + \frac{1}{2}\rho u^2 (\frac{1}{A_1^2} - \frac{1}{A_2^2}) \qquad \text{Eq. (S2)}$$

Based on the pressure in the tubing just before entering the flow cell, $P_2$, we calculate the pressure inside the flow cell, $P_3$ by again applying Bernoulli's Law. However, we first make the assumption that the fluid motion inside the hybriwell is laminar and without head loss, and that $v_3$ inside the flow cell is $v_3 = \frac{u}{A_3}$. This gives:

$$P_3 = P_2 + \rho g h_1 + \frac{1}{2}\rho u^2 (\frac{1}{A_2^2} - \frac{1}{A_3^2}) \qquad \text{Eq. (S3)}$$

Substituting Eq. S2 into Eq. S3 finally gives:

$$P_3 = P_1 + \rho g h_1 + \frac{1}{2}\rho u^2 (\frac{1}{A_1^2} - \frac{1}{A_3^2}) \qquad \text{Eq. (S4)}$$

$P_1$ is calculated using data from the syringe pump user manual (NE-1000, New Era Pump Systems Inc), which indicates that the minimum and maximum pressures applied on a 1 mL syringe are 13 psi and 667 psi at flow rates of 0.0121 µL/min and 881 µL/min respectively. We assume a linear relation between the syringe pump pressure and flow rate, from which we can interpolate syringe pump pressures. $P_3$ is calculated using the parameters in the 'Value' column of Table S2, estimating the pressure inside the hybriwells at varying flow rates.

**s13. (fig. S11) Refractive Index Matching Does Not Change Visible Analyte Distribution**

We performed super-resolution imaging measurements of rhodamine 6G analyte dissolved in media of matching refractive index to silica and cellulose. A 1 nM rhodamine 6G (99%, Fisher) index-matched solution was prepared by dissolving 60% wt. D-(+)-glucose (>99.5%, Sigma Life Sciences) solid powder in water. Glucose powder was added in small increments to a water solution heated to 50° C, while stirring. The final solution was vortexed and allowed to cool to room temperature before adding solid rhodamine 6G, then the mixture was sonicated and further vortexed to aid dissolution into the viscous solution.

Glucose solutions have been shown to change in refractive index with concentration (*57*), approximately matching the 1.47-1.48 index of cellulose and silica under 532 nm illumination at a ~60+% wt. concentration. Here we only went up to 60% as past that the solution becomes too viscous to effectively flow through our system. Single-molecule imaging was then carried out as in the rest of the text, focusing on untreated Cellulose-B FPPs. The resulting super-resolution maps (fig. S11A) show the same spatial distribution as those without index matching media, with little to no analytes reaching the inner volume of the particles. Similarly, the quantified cumulative analyte distributions show no significant change compared to the results done in non-index-matched HEPES buffer solutions (fig. S11B). This – along with the fact that analytes are visible and identified within bare silica and Whelk-O1 FPPs – supports that we can reliably image analytes within the volume of these chromatography materials.

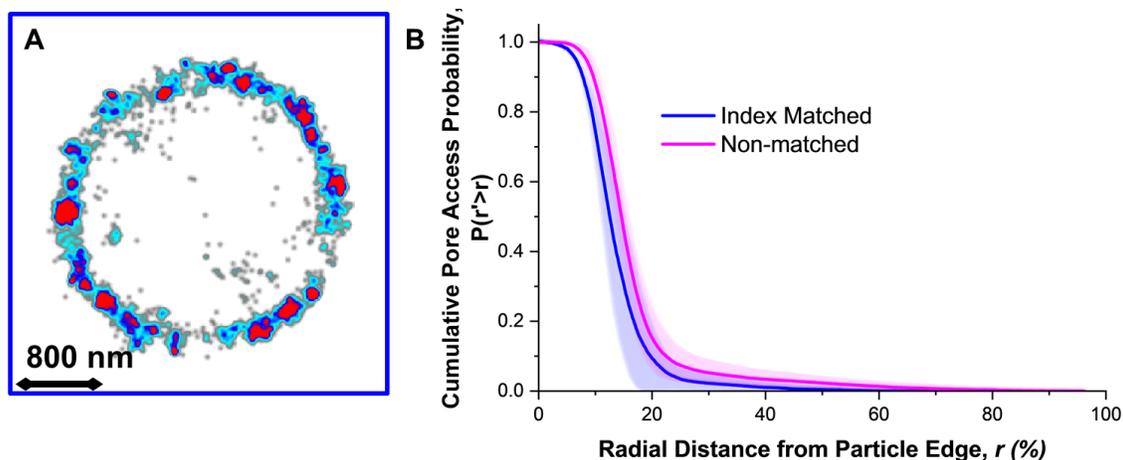

**fig. S11. Spatial distribution of analytes within the volume of particles is unchanged with refractive index matching.** (**A**) The super-resolution map of imaged isolated porous Cellulose-B particle shows that analytes fail to enter the inner volume of particles. (**B**) The cumulative distribution of analytes under refractive index-matched and non-matched conditions are identical, demonstrating that analytes are not being obscured by the cellulose functionalization.

**s14. DMSO Treatment for Removal of Functionalized Cellulose**

Cellulose-B particles consist of a porous silica ($SiO_2$) matrix that is coated is with cellulose, with functional end groups (*tris(3,5 dimethyl phenyl-carbamate)*). The "fully-porous" stationary phase particles were treated with different concentrations of Dimethyl Sulfoxide (DMSO; ≥99% pure, MP Biomedicals) dissolved in water. Cellulose-B particles were mixed as 0.1% wt suspensions by vortexing in solutions of 10%, 20%, and 100% v/v DMSO (in $H_2O$). The mixtures were heated on a hot-plate to ~40°C, stirring at 250 rpm, and left as such overnight for ~12 hours. Then, the solutions were removed from the hot-plate and centrifuged at 4000 rpm for 10 minutes. The supernatant solution was removed until only the solids at the bottom of the containers (20 mL glass vials in this case) remained. The vials were then refilled with water and vortexed, then stirred at 250 rpm ~5 min. This was followed by another round of centrifuging, removing the supernatant, and another wash (by filling with water and stirring), and repeated at least three times total per solution. Finally, the solutions with treated particles were either used for microscopy by drop-cast liquid sample deposition, or by centrifuging, removing supernatant, and drying at ~40°C for ~5 hrs to isolate as dry powder for other experiments. Cellulose-B columns for HPLC measurements followed different DMSO treatment procedure, detailed section s19.

**s15. (fig. S12) 3D Mapping of DMSO (100%) Treated Particle**

3D mapping of Cellulose-B particle after treatment with 100% DMSO solution for 12 hr. Imaging done under the same procedure and conditions as other particles in Figure 2 in the main text. 3D slices of the map are included to provide a clearer view of the interior distribution of analytes.

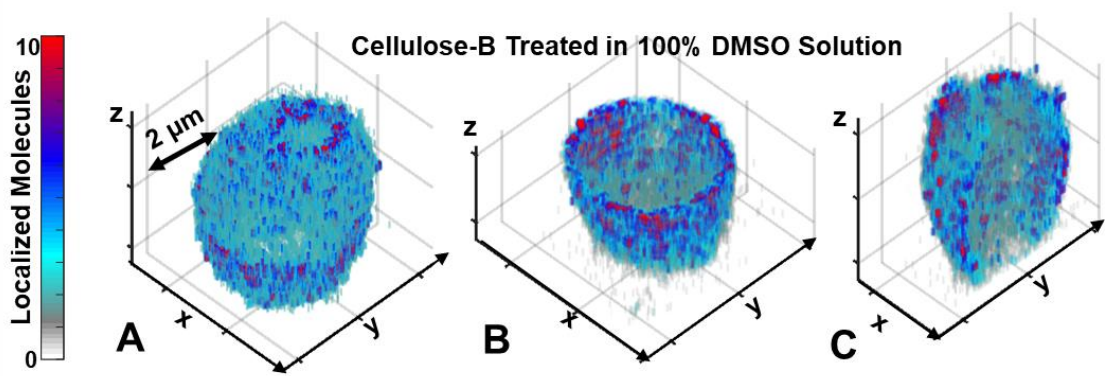

**fig. S12. Single-molecule fluorescence characterization of chromatography particles shows regained pore accessibility after treatment with organic solvent.** (**A**) The 3D super-resolution map of a rhodamine 6G single molecule adsorptions measured throughout the volume of a Cellulose-B porous chromatography particle. The particle was treated with 100% DMSO solution at 40°C over 12 hrs to remove outer layer of functional groups, regaining accessibility to the inner volume of the particle. (**B**) A horizontal cross section of the particle at the half-height shows visible analytes present within the particle volume. Ring-like features are due to the 250 ± 10 nm spacing between the collected imaging areas (slices). (**C**) Vertical cross section further shows that analyte distribution within the particle volume remains consistent throughout.

**s16. (fig. S13) ToF-SIMS Elemental Mapping Confirms the Presence and Removal of Dense Functionalized Groups in the Inner Volume of the Porous Particles.**

We used the elemental mapping capability of Time of Flight - Secondary Ion Mass Spectrometry (ToF-SIMS) to corroborate our super-resolution microscopy results and further reveal that both the cellulose and chlorine-containing tris(3,5 dimethyl phenyl carbamate) are reduced with DMSO treatment (fig. S13). We sought to confirm the spatial distribution of the functional cellulose coating within the chromatography particles, which were imaged after cross sectioning via ion beam milling. Unpacked 5 μm diameter, 1000 Å pore, Cellulose-B (Regis) particles were immobilized as a conductive silver paste (Ted Pella, PELCO Conductive Silver 187). A small drop of conductive silver paint was first deposited on a stainless steel planar milling blade; the dry sample powder was pressed into it, and then lightly tapped vertically to allow excess to fall off. Ion milling was done on a Gatan Model 693 Ilion+ Precision Cross-Section Ion Milling System, which includes a PIPS Cold Stage Controller. The blade was loaded into the system and pumped down to 3.0 x10$^{-5}$ Torr and allowed to cool (via a liquid nitrogen PIPS cold stage) to -40 °C for 20 min. The sample was then ion milled (UHP Ar background) for 60 min at an accelerating voltage of 5.5 keV. Following milling, the system was vented and the sample removed.

Elemental mapping (fig. S13 overlays) was carried out on a Physical Electronics nanoTOF TRIFT V system, with a Primary Liquid Metal Ion Gun (LMIG) set to $^{69}$Ga$^{+}$, 30 kV, 1 nA (DC). Acquisition was carried out in unbunched negative mode, with e-gun charge compensation, aperture size of 50 μm, and an image scan size of 30 μm. Sample surface was cleaned by light sputtering using a 3 kV, 1000 nA, Ar gas gun, set on a 100 μm x 100 μm area for 15 seconds. Mapping was conducted before and after sputtering to confirm sample stability (immobilization) and cleanliness, with before and after images showing little to no change after sputtering. Ion count signal was collected for ~30 minutes at each sample area, until reaching ~7x10$^{6}$ total ion counts. The secondary electron detector (SED) option on the ToF-SIMS instrument was also used to capture secondary electron images purely as a visual reference of sample morphology (fig. S13 black and white). Elemental mapping of cross-sectioned particles before (fig. S13A) and after DMSO treatment (fig. S13B) was done with $^{35}$[Cl]$^{-}$, $^{12}$[C]$^{-}$, $^{28}$[Si]$^{-}$, and $^{16}$[O]$^{-}$ being our primary reference peaks for monitoring Cl, C, Si, and O ion contents; however, for each measurement, mass range up to 1850 amu was acquired.

ToF-SIMS elemental mapping confirmed the removal of functional groups, and the presence of an outer "shell" that leads to pore blocking in untreated Cellulose-B particles. The accumulation of chlorine-containing groups at the outer surface of particles (fig. S13A, overlay) is indicative of the formation of an outer "shell" of functionalized material that is the likely cause of pore blocking. This shell was then removed by the DMSO solvent treatment (fig. S13B, overlay), leaving a more homogenous distribution of functional groups, while allowing analytes and solvent to penetrate into the porous particles. We observe a drop from 75,966 $^{35}$[Cl]$^{-}$ peak counts (normalized counts: 0.01084) down to 23,725 (normalized counts: 0.00338) after DMSO treatment, indicating a ~70% removal of the visible functional groups.

We finally note that the uneven particle surface deposition and the harsh ion beam milling result in a large difference between the detectable distribution of Cl counts, given different depth of sectioning, and distances from the detector. This is further complicated by the small amount of chlorine groups present in the sample to begin with, which are difficult to map out without long collection times. Regardless, we were able to observe a clear decrease in Cl groups after

DMSO treatment, as well as a shift in the overall distribution, where the high density of functional groups near the outer edge of the particles was removed.

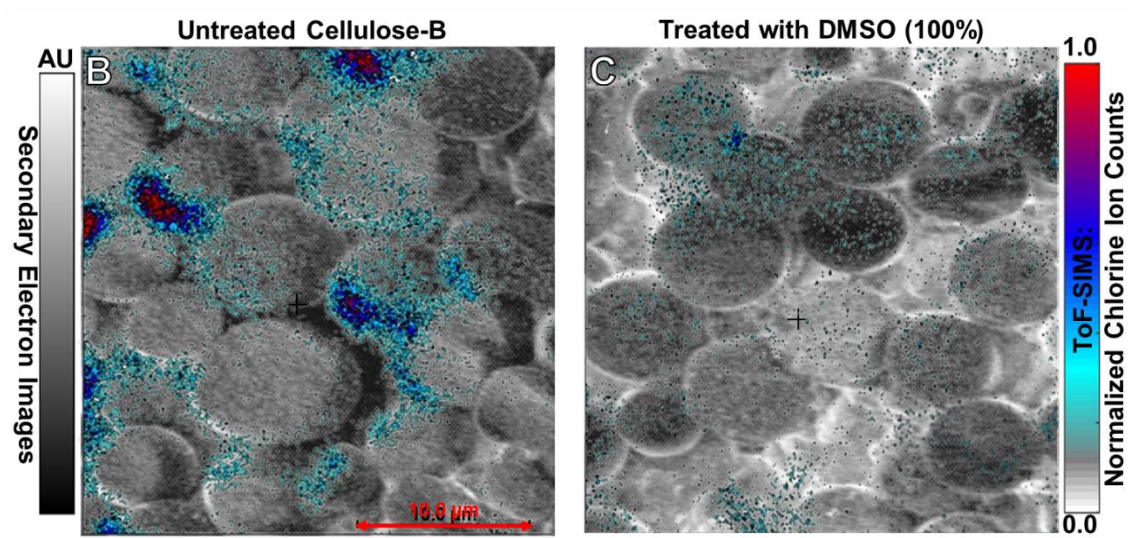

**fig. S13. ToF-SIMs elemental mapping of chromatography particle cross-sections before and after treatment with DMSO solvent shows removal of an outer functionalized shell, and reduced degree of functionalization.** (**A**) The accumulation of chlorine-containing groups at the outer surface of particles (overlay) is indicative of the formation of an outer "shell" of functionalized material that is the likely cause of pore blocking. (**B**) This shell is removed by the DMSO solvent treatment (overlay), leaving less dense, but more homogenous distribution of functional groups. Chromatography particles were cross sectioned by immobilizing on conductive silver paste then ion milling for 60 minutes at an accelerating voltage of 5.5 keV

**s17. (fig. S14) Free Energy Derived from Adsorption Kinetics**

The free energy of adsorption can be experimentally estimated from the cumulative dwell and association time distribution of single-molecule analyte observations. The measured dwell time distribution for sorption sites is fitted to a single-term decaying exponential for weakly adsorbing sites, and a two-term decaying exponential for strong specific adsorption (*58*).

$$P(t' > t) = C_1 e^{-t \cdot k_1} + C_2 e^{-t \cdot k_2} \qquad \text{Eq. (S5)}$$

The rate constants ($k_1$, $k_2$) of the exponentials correspond to desorption rates for weak and strong specific adsorption respectively. The same can then be applied to the association time distribution data, giving rates for adsorption. These can then be converted to free energy (*ΔG*) terms by the equation:

$$\Delta G = -RT ln(k_{eq}) \qquad \text{Eq. (S6)}$$

where *R* is the gas constant, and *T* is the temperature (in Kelvin) (*29*) and $k_{eq}$ is the ratio of the adsorption ($k_a$) over the desorption ($k_d$) rate constant, obtained from the association and dwell time distributions, respectively. Here we took *T* to be 20 °C (293K). Following this procedure, we get a minimum measured free energy of -2.9 ± 0.9 kJ/mol for specific adsorption, and -11.4 ± 0.6 kJ/mol maximum for non-specific adsorption, consistent with hydrogen bonding, hydrophobic, or electrostatic interactions between the analyte and stationary phase.

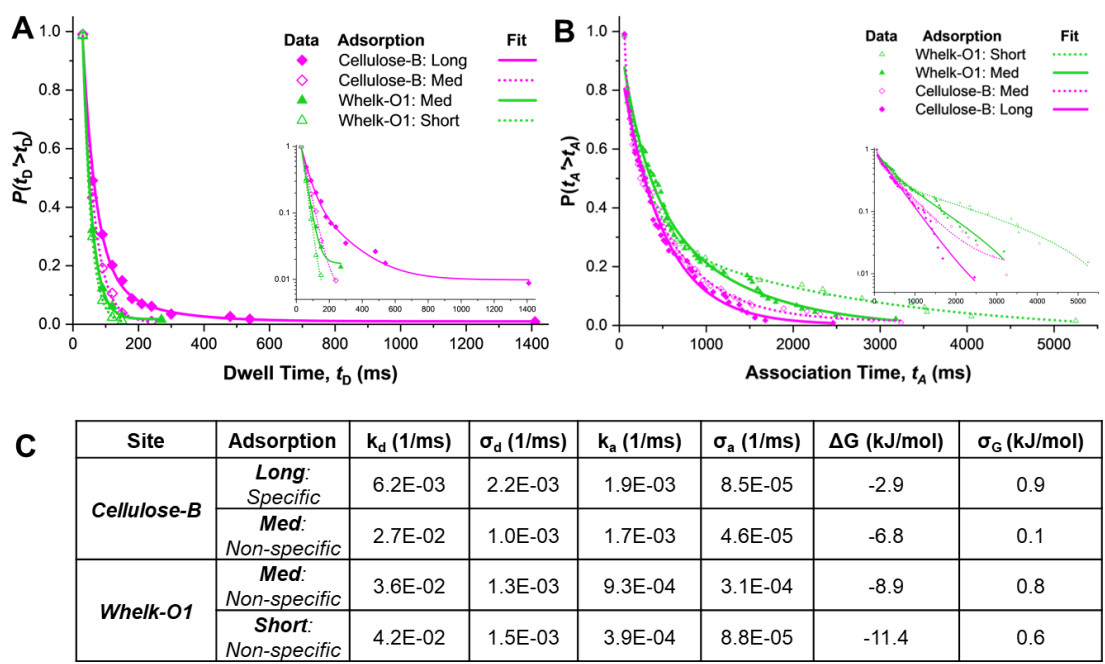

| C | Site | Adsorption | $k_d$ (1/ms) | $\sigma_d$ (1/ms) | $k_a$ (1/ms) | $\sigma_a$ (1/ms) | ΔG (kJ/mol) | $\sigma_G$ (kJ/mol) |
|---|---|---|---|---|---|---|---|---|
| | **Cellulose-B** | **Long**: Specific | 6.2E-03 | 2.2E-03 | 1.9E-03 | 8.5E-05 | -2.9 | 0.9 |
| | | **Med**: Non-specific | 2.7E-02 | 1.0E-03 | 1.7E-03 | 4.6E-05 | -6.8 | 0.1 |
| | **Whelk-O1** | **Med**: Non-specific | 3.6E-02 | 1.3E-03 | 9.3E-04 | 3.1E-04 | -8.9 | 0.8 |
| | | **Short**: Non-specific | 4.2E-02 | 1.5E-03 | 3.9E-04 | 8.8E-05 | -11.4 | 0.6 |

**fig. S14. Free energy derived from single-molecule measurements of adsorption kinetics.** (**A**) Experimental kinetic adsorption data (measured as dwell times) are plotted as cumulative distributions. These can be modeled as and fit to two-term decaying exponentials, corresponding to weak and strong specific adsorption respectively. (**B**) Association time distributions are simultaneously obtained from the measured time between adsorptions. Insets show semi-log versions of the plots. (**B**) The results of the fitted exponentials can then be converted to free energy (*ΔG*) by using the adsorption and desorption rates constants obtained from the fits of the time distributions.

**s18. (fig. S15) Lévy Process Model Representation of Elution**

Measured single-molecule adsorption kinetics in porous stationary phase particles were used to model and reveal differences in larger scale chromatographic elution. Elution through a chromatography column is modeled by as a stochastic process based on Lévy stochastic processes description. This approach relies on a "real discrete distribution" of sorption times obtained by single-molecule dynamics observation and is well described in Pasti et. al (*30*), with example application in Kisley et. al (*59*).

Briefly, the stochastic model of elution starts by converting the cumulative (Poisson) distribution of the duration of adsorption/desorption events to the frequency domain ($\omega$), then accounting for the discontinuities by using the Lévy representation. The distribution (*f*) of time spent in the stationary phase (dwell time, $t_D$) is related to characteristic function ($\phi$) formalism as:

$$\phi(t_D, \omega, t_A) = \exp[r_m \sum_{i=1}^{i=k} (e^{i\omega t_{D,i}} - 1) \cdot f(t_{D,i})] \quad \text{Eq. (S4)}$$

where $t_A$ is the time spent in the mobile phase (i.e. association/desorption time), for an analyte that has adsorbed to the stationary phase $r_m$ times, and *k* is the index of discrete set of desorption times. By performing a Fourier transform, Eq. S4 is converted to the time domain, allowing for simulation of the chromatographic elution peak. Dwell times of single molecule adsorption are generated by collecting 1000+ frame movies of dye molecules interacting with the porous materials at set vertical positions at the approximate half-height of the particles. The asymmetry of the elution peaks generated from the Lévy process representation of chromatography adsorption developed by Pasti et al.(*30*) is dependent on the chosen value $r_m$. However, this $r_m$ value is difficult to equate to the measured single-molecule adsorption events that occur in an imaged area, so it is often chosen empirically. Here we observed that this asymmetry decays exponentially with increasing $r_m$, down to a floor of ~1.1 (fig. S15). The minimum asymmetry is approximately reached when $r_m$ is of comparable magnitude to the number observed time steps used in the model. To reduce variability on our model elution curves we selected an $r_m$ value of 1000, at which point the decay curve for asymmetry begins to plateau.

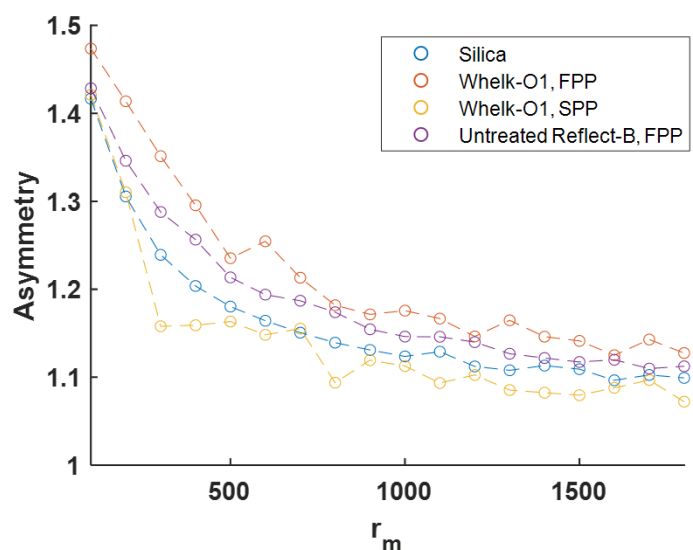

**fig. S15. Levy process model chromatographic peaks decrease in asymmetry with increasing average number of adsorption events.** The asymmetry of the elution peaks generated from the Levy process model representation of chromatography adsorption developed by Pasti et al.(*30*) is dependent on the chosen value $r_m$, which corresponds to the average number of a given molecule adsorbs over the elution through a column. We observe that this asymmetry decays exponentially with increasing $r_m$, down to a floor of 1.1. The minimum asymmetry is approximately reached when $r_m$ is of comparable magnitude to the number observed time steps used in the model, which were 3000 (frames) in this case.

**s19. (fig. S16.) HPLC: Bulk Chromatography Measurements**

All bulk chromatography experiments were performed on a Shimadzu liquid chromatography instrument (LC-20AD Prominence) with photodiode array detection (SPD-M20A, Prominence) with an isocratic mobile phase of 70% ethanol (EtOH, 200 proof, ACS grade from Pharmaco) and 30% HEPES (20 mM, pH 7.33, 2-[4-(2-hydroxyethyl)-1-piperazinyl]-ethane sulfonic acid, high purity from VWR).

A REFLECT C-Cellulose B (5 µm, 5 cm x 4.6 cm from Regis Technologies) column was pre-treated with dimethyl sulfoxide (DMSO, HPLC grade from Fisher) at 30 ºC as follows.  First, 20 vol% DMSO:H2O was flowed through the column at 0.1 mL/min for 10 min, 0.2 mL/min for 10 min, 0.3 mL/min for 3 hrs, and 0.4 mL/min for 10 min; the 20 vol% DMSO:H2O was then left in the column, stagnant, for 18 hours.  This slow increase of flow rate was found necessary due to excessive column backpressure if high (>0.3 mL/min) flow rates were initially employed.  Finally, the column was flushed with 20 vol% DMSO:H2O at 0.3 mL/min for 30 min.  Prior to experiments, the DMSO-treated column was rinsed with mobile phase (70% EtOH:30% HEPES) at 0.1 mL/min for 2 hrs then at 0.3 mL/min for 2 hrs.  The control (untreated) column was rinsed with mobile phase (70% EtOH: 30% HEPES) at 0.3 mL/min for at least 2 hrs prior to experiments.

Rhodamine 6G (3 µL, 100 µM in HEPES buffer, from Biotang Inc) was run on each column with an isocratic mobile phase of 70% EtOH: 30% HEPES, a flow rate of 0.5 mL/min, and a column temperature of 30 ºC.  Detection was accomplished using a photodiode array over the range 200 – 600 nm at 5 Hz.  The resulting chromatograms were visualized in Shimadzu Lab Solutions software, extracted at 530 nm (4 nm bandwidth), exported, and further analyzed with Origin 2022.

Comparison to the model elution curves generated from single-molecule measurements show remarkably similar peak shapes, with comparable tailing (fig. S16). Both the model and HPLC measured elutions show asymmetries (measured as the ratio of the width of the right side of the peak over the width of the left side), favoring the right side of the curve (known as tailing), which are associated with strong retention of analyte in the stationary phase. Both also show little to no change in the asymmetry of the curves before and after treatment of Cellulose-B particles with DMSO, supporting our conclusions that the particles remain sufficiently functionalized to achieve a separation even after treatment. However, the degree of tailing differs between the model and HPLC measurements, where we find that the experimental bulk asymmetry (up to a degree of 2) can be up to 1.75x what the model predicts (up to a degree of 1.5). Similarly, the change in broadening is lower for the model data. Where the model shows a 13% decrease in peak width, the HPLC results show broadening with an increase of 50% peak width. We can attribute these differences to the limitation of the single molecule model, where it only accounts for adsorption mass transfer behavior, while the elution peak in a column is influenced by many other factors such as convection, Eddy, and longitudinal diffusion due to column length and packing.

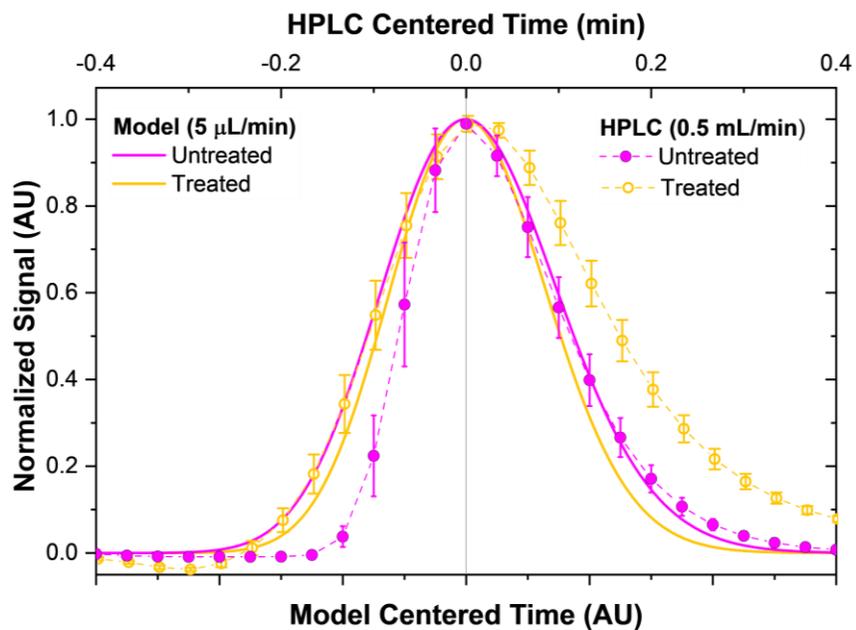

**fig. S16.** HPLC system elution measurements with packed Cellulose-B columns before and after treatment with DMSO solvent show similar peak shape and tailing as those generated with our single-molecule model of elution curves.